\documentclass[11pt,a4paper]{article}
\usepackage[pdftex]{graphicx}%[dvipdfmx]{graphicx}%%% Include figure files
\usepackage{mathptmx}
\usepackage{amsmath,amssymb}
\usepackage{times}
\usepackage[T1]{fontenc}
\normalfont
\usepackage{fourier} 
\usepackage{color}
\newcommand{\ie}{{\it i.e.}\,}
\newcommand{\eg}{{\it e.g.}\,}

\begin{document}
\renewcommand{\thefootnote}{\#\arabic{footnote}} 
\setcounter{footnote}{0}

%%%%%%%%%%%%%%%%%%%%%%%%%%%%%%%%%%%%%%%%%%%%%%%%%%%%%%%%%%%%%%%%%%%%%%
%%%%%  Title Page  %%%%%%%%%%%%%%%%%%%%%%%%%%%%%%%%%%%%%%%%%%%%%%%%%%%
\begin{center}

%%%%%%%% Preprint #
\begin{flushright}
\end{flushright}

\begin{Large}
  {\bf Resonant leptogenesis at TeV-scale\\[0.5ex]
    and neutrinoless double beta decay}
\end{Large}

\vspace{1cm}
Takehiko Asaka and Takahiro Yoshida$^\dagger$

\vspace{0.5cm}

{\em \small Department of Physics, Niigata University, Niigata 950-2181, Japan}
\\[1ex]
$^{\dagger}$
{\em \small
Graduate School of Science and Technology, Niigata University, Niigata 950-2181, Japan}
\end{center}

\vspace{0.5cm}

\begin{center}
%(\today)
(December 29, 2018)
\end{center}

\vspace{1cm}

\begin{abstract}
  We investigate a resonant leptogenesis scenario by quasi-degenerate
  right-handed neutrinos which have 
  TeV-scale masses.
  Especially, we consider the case when two right-handed neutrinos
  are responsible to leptogenesis and the seesaw mechanism
  for active neutrino masses, and assume that 
  the CP violation occurs only in the mixing matrix of active neutrinos.
  In this case the sign of the baryon asymmetry depends
  on the Dirac and Majorana CP phases
  as well as the mixing angle of the right-handed neutrinos.
  It is shown how the yield of the baryon asymmetry correlates
  with these parameters.  In addition, we find that
  the effective neutrino mass in the neutrinoless double beta decay
  receives an additional constraint in order to account for
  the observed baryon asymmetry depending on the masses and mixing angle of right-handed neutrinos. 
\end{abstract}

\vspace{4cm}
%\pacs{ 
%PACS numbers
%\hspace{0.5cm} 
%13.15.+g, %% Neutrino interactions  
%14.60.Pq, %% Neutrino mass and mixing 
% }
%Keywords
\hspace{0.5cm}

\newpage
%%%%%%%%%%%%%%%%%%%%%%%%%%%%%%%%%%%%%%%%%%%%%%%%%%%%%%%%%%%%%%%%%%%%%%%
\section{Introduction}
%%%%%%%%%%%%%%%%%%%%%%%%%%%%%%%%%%%%%%%%%%%%%%%%%%%%%%%%%%%%%%%%%%%%%%%
Leptogenesis~\cite{Fukugita:1986hr} 
is an attractive mechanism accounting for the baryon asymmetry 
of the Universe (BAU).  
See, for example, reviews~\cite{Buchmuller:2005eh,Davidson:2008bu}.
In the canonical scenario 
the out-of-equilibrium decays of right-handed neutrinos, $\nu_R$'s, generate a lepton asymmetry, which is partially converted into the baryon asymmetry through the
sphaleron effect at high temperatures \cite{Kuzmin:1985mm}.   When their masses are hierarchical,
the observed BAU~\cite{Aghanim:2018eyx}
\begin{align}
	\label{eq:BAU}
	\left. Y_B \right.^{\rm OBS} = 
	\left. \frac{n_B}{s} \right|_{\rm obs}
	= (0.870 \pm 0.006) \times 10^{-10} \,,
\end{align}
where $Y_B$ is the ratio between the baryon number density $n_B$ and 
entropy density $s$, 
can be explained if their masses are heavier than ${\cal O}(10^9)$~GeV~\cite{Giudice:2003jh}.%
\footnote{
The lower bound on masses can be reduced as ${\cal O}(10^6)$~GeV
if one considers the non-thermal production of right-handed neutrinos
via the inflaton decays~\cite{Leptogenesis_inflation_decay}.
}  

It should be noted that such superheavy particles 
can also give a significant impact on neutrino masses.
The various oscillation experiments have shown that neutrinos have very suppressed but non-zero masses.
The smallness of the masses can be naturally explained by
the seesaw mechanism with superheavy $\nu_R$'s~\cite{Seesaw}.  

Leptogenesis can operate even if $\nu_R$'s masses are much smaller than the above value, which is resonant leptogenesis~\cite{Pilaftsis:2003gt}.
The mass degeneracy of $\nu_R$'s enhances the CP violating effects,
which leads to the resonant production of lepton asymmetry by their decay.%
\footnote{
A sufficient amount of the BAU can be generated 
by right-handed neutrinos even with masses
${\cal O}(1)$~MeV-${\cal O}(10^2)$~GeV 
if one uses the flavor oscillations of $\nu_R$'s~\cite{Akhmedov:1998qx,Asaka:2005pn,Canetti:2010aw,Asaka:2013jfa}.
}
The required mass degeneracy may be a consequence of the symmetry of the model.

In resonant leptogenesis scenario, since it can occur
at relatively lower temperatures,
the flavor effects of leptogenesis~\cite{Abada:2006fw,Nardi:2006fx,Abada:2006ea,Blanchet:2006be,Pascoli:2006ie,Pascoli:2006ci,Moffat:2018smo,DeSimone:2006nrs}~can be essential.
In such a case, the yield of the BAU depends on 
the mixing matrix of active neutrinos $U$ called as the Pontecorvo-Maki-Nakagawa-Sakata (PMNS) matrix~\cite{PMNS}, 
and hence the CP violation due to
the Dirac and/or Majorana phases in $U$
can be an origin of the BAU.

We consider here resonant leptogenesis by 
right-handed neutrinos with TeV-scale masses in the framework of the seesaw mechanism.
Especially, it is investigated 
the case in which the CP violation occurs only 
in the mixing matrix $U$ of active neutrinos.
We will show that how the yield of the BAU depends on 
the Dirac and/or Majorana phases.

Majorana masses of $\nu_R$'s break the lepton number
by two units, which is necessary for leptogenesis.  Then, various processes, which are
absent in the Standard Model, are predicted by the models with the seesaw mechanism. One important
example is the neutrinoless double beta ($0 \nu \beta \beta$) decay
$(Z, A) \to ( Z+ 2, A) + 2 e^{-}$~\cite{Pas:2015eia}.
The rates of such decays are parameterized by the so-called
effective mass of neutrinos $m_{\rm eff}$, which depends on the CP violating parameters of active neutrinos.
We then discuss the possible relation between the yield of BAU
and $m_{\rm eff}$.

The present article is organized as follows:
In Section 2 we explain the framework of the analysis in which 
the properties of right-handed neutrinos are specified.
In Section 3 we present the method of estimating the BAU
through resonant leptogenesis including the flavor effects. We then show how $Y_B$ depends on the CP violating parameters of active neutrinos.
It is discussed in Section 4 that how the conditions
 accounting for the observed BAU give the impacts on the 0$\nu \beta \beta$ decay.
We will show that the BAU provides the upper and/or lower bound
of the effective mass $m_{\rm eff}$ in some cases,
especially when the mass difference of $\nu_{R}$'s becomes larger.
The final Section is devoted to conclusions.
We add Appendix A to present the Boltzmann equations 
used in the analysis.

%%%%%%%%%%%%%%%%%%%%%%%%%%%%%%%%%%%%%%%%%%%%%%%%%%%%%%%%%%%%%%%%%%%%%%
\section{TeV-scale right-handed neutrinos}
%%%%%%%%%%%%%%%%%%%%%%%%%%%%%%%%%%%%%%%%%%%%%%%%%%%%%%%%%%%%%%%%%%%%%%
First of all, let us explain the framework of the present analysis.
We consider the Standard Model (SM) extended by three right-handed neutrinos 
$\nu_{R \, I}$ ($I=1,2,3$) with the Lagrangian
\begin{align}
	{\cal L} 
	&=
	{\cal L}_{\rm SM}
	+ i \, \overline{\nu}_{R I} \gamma^\mu \partial_\mu 
	\nu_{R I} - 
	\left( \,
		F_{\alpha I} \, \overline{\ell}_{\alpha} H \nu_{R \, I}
		+
		\frac{[M_M]_{IJ}}{2} \, \overline{\nu}_{R \, I}^c \nu_{R \, J}
		+
		h.c. \,
	\right) \,,
\end{align}
where ${\cal L}_{\rm SM}$ is the SM Lagrangian,
and 
$\ell_\alpha$ ($\alpha = e, \mu, \tau$) and $H$ 
are lepton and Higgs doublets, respectively.
$F_{\alpha I}$ are Yukawa coupling constants and
$M_M$ are Majorana mass matrix of right-handed neutrinos.
We take the basis in which 
the charged lepton mass matrix and $M_M$ are both diagonal,
and we write $[M_M]_{II} = M_I$ (We take $M_I$ is real and positive.).

We apply the seesaw mechanism for generating the suppressed masses 
of active neutrinos, and work in the parameter range 
$M_I \gg |[M_D]_{\alpha I}| =|F_{\alpha I}|\langle H \rangle$.
In this case, the lighter mass eigenstates are
active neutrinos $\nu_i$ ($i=1,2,3$) and their masses 
$m_i$ are found from the seesaw mass matrix
$M_\nu = - M_D^T M_M^{-1} M_D$
which is diagonalized as $U^\dagger M_\nu U^\ast = D_\nu =\mbox{diag} 
(m_1, m_2, m_3)$.  
Here $U$ is the PMNS matrix~\cite{PMNS}, which is represented as
\begin{align}
	U = 
  \left( 
    \begin{array}{c c c}
      c_{12} c_{13} &
      s_{12} c_{13} &
      s_{13} e^{- i \delta_{\rm CP}} 
      \\
      - c_{23} s_{12} - s_{23} c_{12} s_{13} e^{i \delta_{\rm CP}} &
      c_{23} c_{12} - s_{23} s_{12} s_{13} e^{i \delta_{\rm CP}} &
      s_{23} c_{13} 
      \\
      s_{23} s_{12} - c_{23} c_{12} s_{13} e^{i \delta_{\rm CP}} &
      - s_{23} c_{12} - c_{23} s_{12} s_{13} e^{i \delta_{\rm CP}} &
      c_{23} c_{13}
    \end{array}
  \right)  
  \times
  \mbox{diag} 
  ( 1\,,~ e^{i \alpha_{21}/2} \,,~ e^{i \alpha_{31}/2} ) \,,
\end{align}
where $s_{ij} = \sin \theta_{ij}$ and $c_{ij} = \cos \theta_{ij}$
with mixing angles $\theta_{ij}$ of active neutrinos.   
$\delta_{\rm CP}$ is the Dirac phase and $\alpha_{21,31}$
are the Majorana phases for CP violation.
The global analysis of three flavor neutrino oscillations \cite{Esteban:2016qun}
provides the mixing angles and the mass squared differences 
$\Delta m_{ij}^2 = m_i^2 - m_j^2$
as shown in Table~\ref{tab:NuFIT}.
Notice that there are two possibilities of mass ordering,
the normal hierarchy (NH) $m_3 > m_2 > m_1$ 
and the inverted hierarchy (IH) $m_2 > m_1 > m_3$.

%%%%%%%%%%
\begin{table}[h]
\begin{center}
\begin{tabular}{| c || c | c | c | c | c |}
\hline
 & 
 $\theta_{12}$ & $\theta_{23}$ & $\theta_{13}$ &
 $\Delta m_{21}^2$~[eV$^2$] & $\Delta m_{3 \ell}^2$~[eV$^2$] 
 \\ \hline
NH & 
	 $33.62^\circ$ & 47.2$^\circ$ & 8.54$^\circ$
  & $7.40 \times 10^{-5}$ 
	& $+ 2.494 \times 10^{-3}$ ($\ell = 1$)  
	\\ \hline
IH & 
	$33.62^\circ$ & 48.1$^\circ$ & 8.58$^\circ$	
	& $7.40 \times 10^{-5}$ 
	& $- 2.465 \times 10^{-3}$ ($\ell = 2$)  
\\ \hline
\end{tabular}
\caption{The mixing angles and the mass squared differences
from the global analysis of three flavor neutrino 
oscillations~\cite{Esteban:2016qun}.}\label{tab:NuFIT}
\end{center}
\end{table}
%%%%%%%%%%

On the other hand, the heavier ones are heavy neutral leptons $N_I$ with
masses $M_I$. Through the non-zero vacuum-expectation value of the Higgs field, they mix with active ones as
$\nu_{L \, \alpha} = U_{\alpha i} \nu_i + \Theta_{\alpha I} N_I$
with $\Theta_{\alpha I} = [M_D]_{\alpha I} M_I^{-1}$.
This mixing induces the weak interaction of heavy neutral leptons
suppressed by $\Theta_{\alpha I}$.
The properties of heavy neutral leptons are then determined
by $M_I$ and $F_{\alpha I}$.  In order to induce the mixing angles
and masses of neutrino oscillation observations,
the Yukawa coupling can be expressed as~\cite{Casas:2001sr}
\begin{align}
	F = \frac{i}{\langle H \rangle}
	U D_\nu^{1/2} \Omega M_M^{1/2} \,,
\end{align} 
where $\Omega$ is the arbitrary $3 \times 3$ 
complex orthogonal matrix, which is parameterized as
\begin{align}
	\Omega =
	\left( 
		\begin{array}{c c c}
			\cos \omega_{12} & \sin \omega_{12} & 0 \\
			- \sin \omega_{12} & \cos \omega_{12} & 0 \\
			0 & 0 & 1 
		\end{array}
	\right)
		\left( 
		\begin{array}{c c c}
			\cos \omega_{13} & 0 &  \sin \omega_{13} \\
			0 & 1 & 0 \\
			- \sin \omega_{13} & 0 &  \cos \omega_{12}
		\end{array}
	\right)
	\left( 
		\begin{array}{c c c}
			1 & 0 & 0 \\
			0 & \cos \omega_{23} & \sin \omega_{23} \\
			0 & - \sin \omega_{23} & \cos \omega_{23} 
		\end{array}
	\right) \,,
\end{align}
where $\omega_{IJ}$ are complex mixing parameters.

In this analysis we consider the case when
only two right-handed neutrinos are responsible to the seesaw mechanism 
as well as leptogenesis for simplicity.  The mass and Yukawa couplings
of the rest right-handed neutrino are taken to be sufficiently heavy
and small, respectively.  
For the NH case 
with $m_3 > m_2 > m_1 = 0$
we consider the two right-handed neutrinos
$\nu_{R2}$ and $\nu_{R3}$ and mixing parameters are taken to be
\begin{align}
	&\omega_{12} = \omega_{13} = 0 \,,~~~ \omega_{23} \neq 0 \,.
\end{align}
On the other hand, for the IH case
with $m_2 > m_1 > m_3 = 0$ we consider 
$\nu_{R1}$ and $\nu_{R2}$ and
\begin{align}
	&\omega_{13} = \omega_{23} = 0 \,,~~~ \omega_{12} \neq 0 \,.
\end{align}

There are three CP violating parameters which can be a source
of the BAU under this situation. They are the Dirac phase $\delta_{\rm CP}$, 
one combination of Majorana phases, \ie, 
$(\alpha_{21}-\alpha_{31})$ or $\alpha_{21}$ for the NH or IH case, 
and the imaginary part of the mixing parameter, \ie,
Im$\omega_{23}$ or Im$\omega_{12}$ for the NH or IH case.
In the present analysis we assume that the CP violation occurs only 
in the PMNS matrix and all the mixing parameters $\omega_{IJ}$ are
real.  See, for example, Refs.~\cite{Pascoli:2006ie,Pascoli:2006ci,Moffat:2018smo}. It is then discussed whether
two right-handed neutrinos whose masses are TeV-scale can produce 
a sufficient amount of the BAU or not.

%%%%%%%%%%%%%%%%%%%%%%%%%%%%%%%%%%%%%%%%%%%%%%%%%%%%%%%%%%%%%%%%%%%%%%
\section{Baryon asymmetry and CP violations in leptonic sector}
%%%%%%%%%%%%%%%%%%%%%%%%%%%%%%%%%%%%%%%%%%%%%%%%%%%%%%%%%%%%%%%%%%%%%%
In this section we investigate the baryogenesis scenario by TeV-scale 
right-handed neutrinos through resonant leptogenesis~\cite{Pilaftsis:2003gt}. 
The BAU in Eq.~(\ref{eq:BAU}) can be explained if their masses are
quasi-degenerate.  We thus take 
$M_3 = M_N + \Delta M/2$ and $M_2 = M_N - \Delta M/2$ for 
the NH case while $M_2 = M_N + \Delta M/2$ and $M_1 = M_N - \Delta M/2$ for the IH case,
where $M_N \gg \Delta M > 0$.
In this case the generation of the BAU is effective at 
TeV-scale temperatures
and then the flavor effects of leptogenesis \cite{Abada:2006fw,Nardi:2006fx,Abada:2006ea,Blanchet:2006be,Pascoli:2006ie,Pascoli:2006ci,Moffat:2018smo,DeSimone:2006nrs} must be taken into account.  This is crucial to produce the baryon asymmetry 
by the CP violation in the PMNS matrix, since such an effect disappears
for unflavored leptogenesis.
From now on we will estimate the yield of the BAU by TeV-scale right-handed neutrinos
and show how it depends on the low energy CP violating parameters 
in the PMNS matrix.

In this work, we use the Boltzmann equations for estimating the amount of 
the produced baryon asymmetry. 
In the $N_I$ decay the interference between tree and 
one-loop diagrams of vertex and self-energy corrections
induces the lepton asymmetry due to the CP violation in 
the neutrino 
Yukawa coupling constants.
It is characterized by the CP asymmetry parameter $\varepsilon_{\alpha I}$ 
which is defined by
\begin{align}
	\varepsilon _ { \alpha I } 
	&= 
		\frac { 
		\Gamma \left( 
			N _ { I } \to \ell_{ \alpha } + \overline { \Phi } 
			\right) 
		- 
		\Gamma \left( 
			N _ {I } \to \overline { \ell_{ \alpha } }  + \Phi \right) 
		} 
		{ 
		\sum_{\alpha} \Gamma \left( N _ {I } \to \ell _ {\alpha} + \overline { \Phi } 
			\right) 
		+ 
		\sum_{\alpha} \Gamma \left( N _ {I } \to \overline { \ell_{\alpha} } 
		+ \Phi \right) }	 \,,
\end{align}
where 
$\Gamma ( 
			N _ { I } \to \ell_{ \alpha } + \overline { \Phi } 
			)$ is the partial decay width for 
			$ N _ { I } \to \ell_{ \alpha } + \overline { \Phi } $.
Now we consider the case with $\Delta M \ll M_N$, and then 
the contribution from the self-energy correction dominates
over that from the vertex correction.  In this case
$\varepsilon_{\alpha I}$ is given by~\cite{Pilaftsis:2003gt} 
\begin{align}		
	\varepsilon_{\alpha I}	
	&\simeq \frac { 1 } { 8 \pi } \sum_{J \neq I}
	\frac { \operatorname { Im } 
	\left[ F _ { \alpha I } ^ { * } F _ { \alpha J }     
	\left( F ^ { \dagger } F \right) _ { I J } \right] } 
	{ \left( F ^ { \dagger } F \right) _ { I I } } 
	\frac { M _ { I } M _ { J } \left( M _ { I } ^ { 2 } - M _ { J } ^ { 2 }  
	\right) } 
	{ \left( M _ { I } ^ { 2 } - M _ { J } ^ { 2 } \right) ^ { 2 } + A ^ { 2 } }  
	\,.
\end{align}
In this equation $A$ denotes a regulator for the degenerate mass.  The estimation based on the Boltzmann equations becomes worse when the mass difference of right-handed neutrinos becomes very small.  It has, however, been shown in   Refs.~\cite{Garny:2011hg,Iso:2013lba} by using the more precise approach with the Kadanoff-Baym equations that the estimation with the regulator $A= M_I \Gamma_I + M_J \Gamma_J$ (where $\Gamma_I$ is the total decay rate of $N_I$) is the good approximation to estimate the maximal value of the BAU.
It is then found that $|\varepsilon_{\alpha I}|$ takes the maximal value
when the condition $A=|M _ { I } ^ { 2 } - M _ { J } ^ { 2 }|$ is satisfied.
This means that the maximal value of $|\varepsilon_{\alpha I}|$ is 
achieved when the mass difference is $\Delta M = \Delta M_\ast \equiv A/(2M_N)$.
From now on, we call the yield of the BAU with $\Delta M = \Delta M_\ast$
as $Y_B^{\rm MAX}$.

We estimate the yield of the BAU
by using the Bolzmann equations for the yields of $N_I$ ($Y_{N_I}$)
and the charges ($X_\alpha = B/3 - L_\alpha$) associated with 
the baryon number $B$ and the lepton flavor number $L_\alpha$.%
\footnote
{The estimation based on the Kadanoff-Baym equation is found in Refs. \cite{Dev:2014laa, Dev:2014wsa}.}
The explicit equations are presented in Appendix~\ref{boltzmanneq}. 
The initial conditions are $Y_{N_I} = Y_{N_I}^{\rm eq}$ 
and $X_\alpha = 0$, where $Y_{N_I}^{\rm eq}$ is the equilibrium value 
of $Y_{N_I}$.  We then solve the equations from the initial temperature
$T_i \gg M_N$%
\footnote{
We take $T_i/M_N = 100$ for the numerical study.
}
 to the final temperature $T_f = T_{\rm sph}$ 
and calculate the yield of the BAU.  Here $T_{\rm sph}$ is the sphaleron freeze-out 
temperature and $T_{\rm sph} = 131.7$~GeV~\cite{DOnofrio:2014rug} for the observed 
Higgs boson mass.

We take $M_N = 1$~TeV as a representative value
and evaluate the maximal value $Y_B^{\rm MAX}$
by setting the mass difference as $\Delta M = \Delta M_\ast$.
In addition, as explained in the previous section, we consider the case when
the CP violation occurs only in the mixing matrix $U$ of active neutrinos,
\ie, we set ${\rm Im} \omega_{IJ} =0$.  
We take the central values of 
the mixing angles $\theta_{ij}$ and the mass squared differences
$\Delta m_{ij}$ shown in Tab.~\ref{tab:NuFIT} for the sake of simplicity.
Under this situation we investigate how $Y_B$ depends on 
the CP phases $\delta_{CP}$ and $\alpha_{ij}$ and the mixing angle 
${\rm Re} \omega_{IJ}$ of $\nu_R$'s. 

%%%%%%%%%%%%%%%%%%%%%%%%%%%%%%%%%%%%%%%%%%%%%%%%%%%%%%%%%%%%%%%%%%%%%% 
%%%%% ** Figure ** %%%%%%%%%%%%%%%%%%%%%%%%%%%%%%%%%%%%%%%%%%%%%%%%%%%
\begin{figure}[t]
	\begin{center}
    \includegraphics[height=8cm]{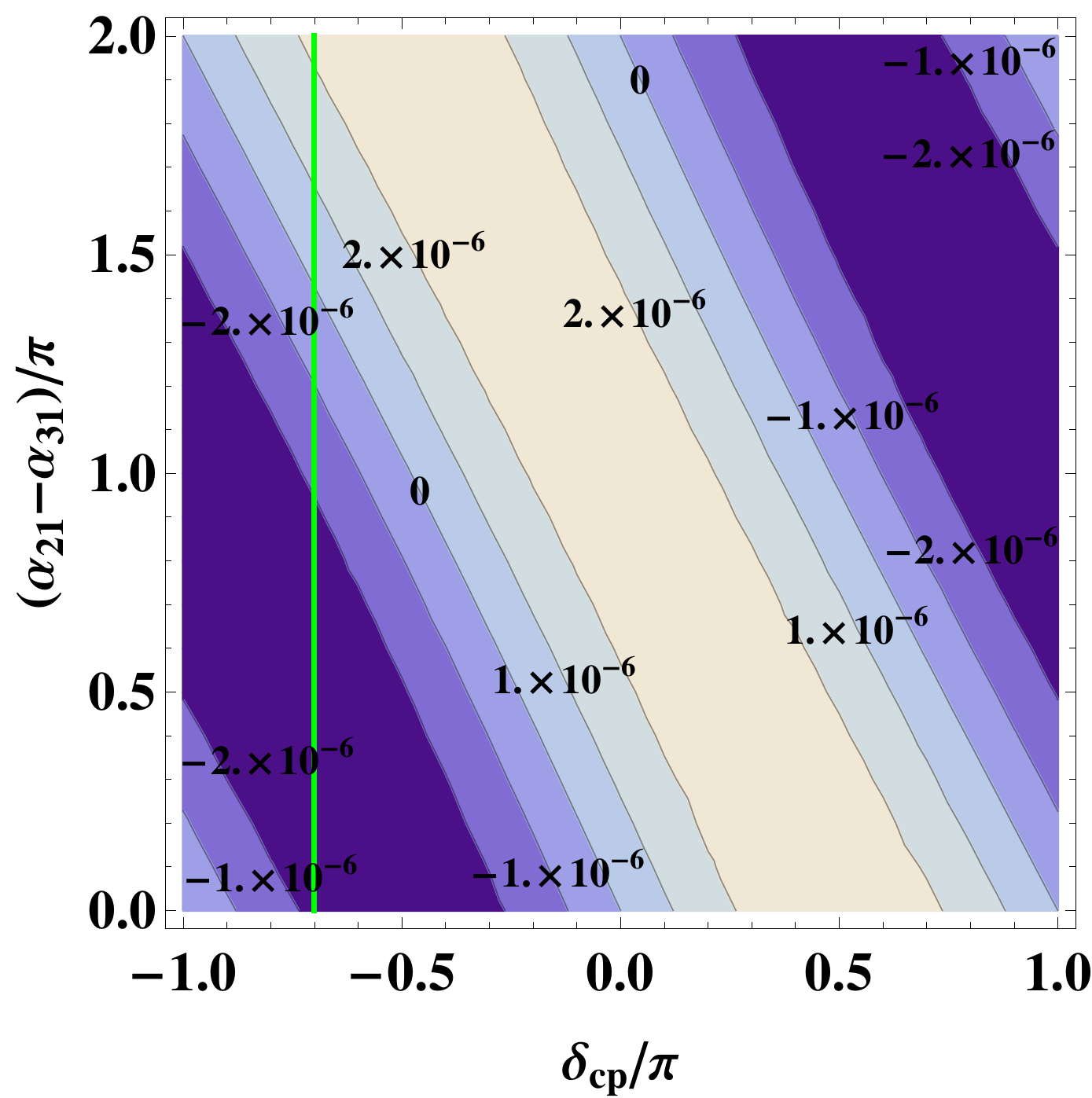}
    \hspace{0.7cm}
    \includegraphics[height=8cm]{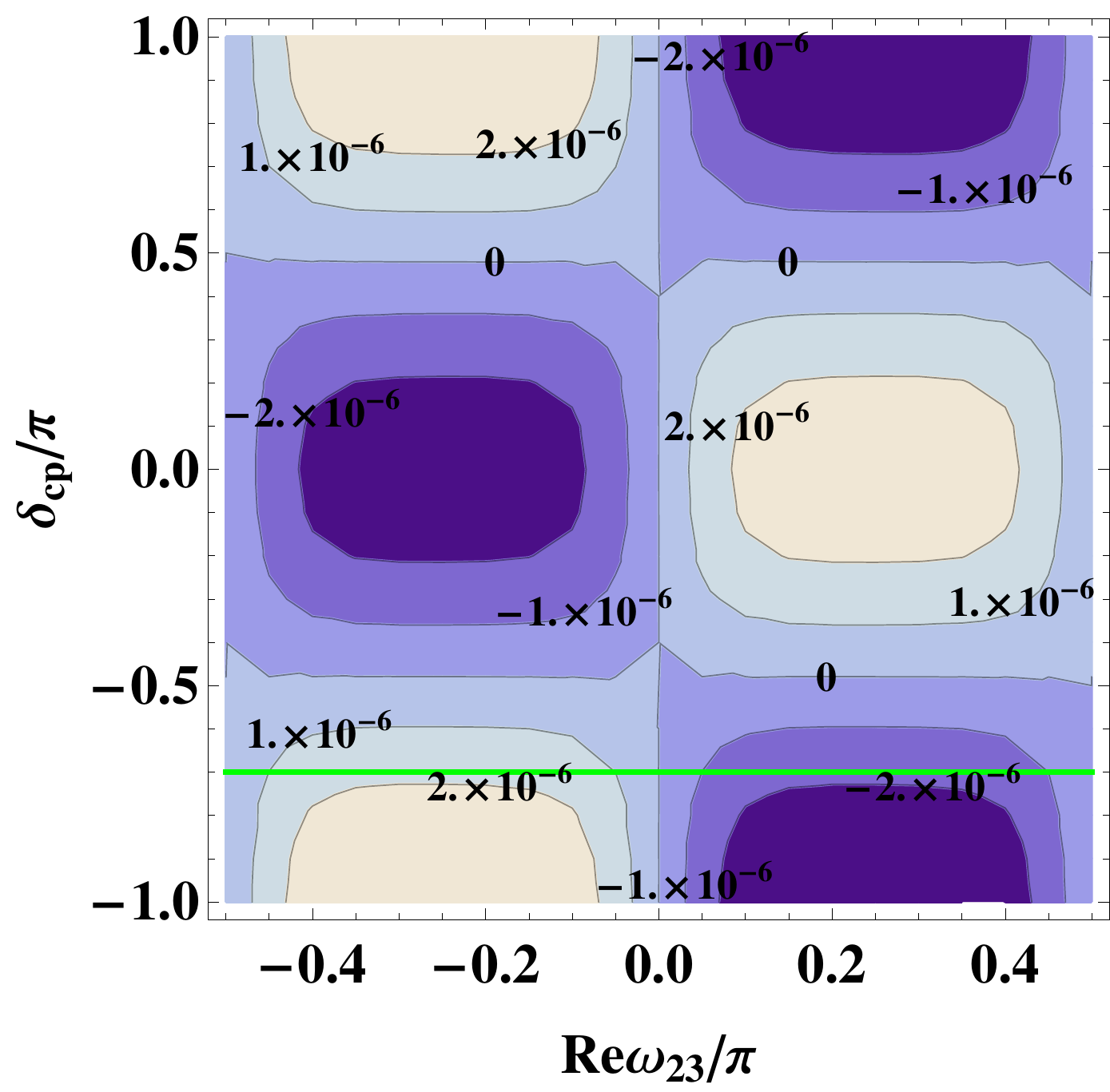}
  \caption{
    Contour plots of $Y_B^{\rm MAX}$ in the NH case.
    $Y_B^{\rm MAX}$ is positive or negative in the region with bright or 
    dark color.
    In the left or right panel the plot is shown in
    the $\delta_{\rm CP}$-$(\alpha_{21}-\alpha_{31})$ plane 
    when Re$\omega_{23} =  \pi/4$ or the Re$\omega_{23}$-$\delta_{\rm CP}$ 
    plane when $\alpha_{21}-\alpha_{31} = \pi$, respectively.
		The green lines shows the central value of $\delta_{\rm CP}$
		from the global neutrino oscillation analysis.
  }
  \label{FIG_YB_NH}
  \end{center}
\end{figure}
%%%%%%%%%%%%%%%%%%%%%%%%%%%%%%%%%%%%%%%%%%%%%%%%%%%%%%%%%%%%%%%%%%%%%%
First, we show the results for the NH case in Fig.~\ref{FIG_YB_NH}.
We find that $Y_B^{\rm MAX}$ can be large as ${\cal O}(10^{-6})$,
which is much larger than the observational BAU in Eq.~(\ref{eq:BAU}).
The left panel represents the contour plot of $Y_B^{\rm MAX}$
in the Dirac and Majorana phase plane by taking 
the mixing angle Re$\omega_{23}=\pi/4$.  
Notice that we have shown the results of $Y_B^{\rm MAX}$ 
by taking $\Delta M = \Delta M_\ast$.  This means that
the observed value $Y_B^{\rm OBS}$ in Eq.~(\ref{eq:BAU}) 
can be explained in the parameter region 
$Y_B^{\rm MAX} \ge Y_B^{\rm OBS}$ by taking $\Delta M \ge \Delta M_\ast$.
The relevant Majorana 
phase is the combination, $\alpha_{21} - \alpha_{31}$, 
in the NH case.
It can be seen that the yield of the BAU does depend on 
both phases significantly. %
\footnote{
The dependence on $\mathrm{Im}\omega_{23}$ is discussed in Ref. \cite{Bambhaniya:2016rbb}.
}
Thus, the experimental information of Dirac phase, \eg,  from accelerator neutrinos~\cite{NOvA:2018gge,Abe:2018wpn}
is crucial for determining the sign of the BAU.
It should be noted that the dependence on the CP violating phases is approximately 
given by
\begin{align}
Y_B \propto \sin \left( \frac{\alpha_{21}-\alpha_{31}}{2}+\delta_{\rm CP} \right) \,,
\end{align}
which is found from the parameter dependence in $\varepsilon_{\alpha I}$
as well as the strength of the wash-out effects, 
\ie\,, the structures in the partial decay rates
$\Gamma \left( N _ {I } \to \ell _ {\alpha} + \overline { \Phi } 
			\right)$.
On the other hand, the right panel in Fig.~\ref{FIG_YB_NH}
shows the contour in the mixing angle Re$\omega_{23}$
and $\delta_{\rm CP}$ plane when $\alpha_{21}-\alpha_{31}= \pi$.
It is found that $Y_B^{\rm MAX}$ depends on Re$\omega_{23}$ 
and the observed BAU cannot be generated when 
the mixing of $\nu_R$'s disappears at Re$\omega_{23}=0$, $\pi/2$.

%%%%%%%%%%%%%%%%%%%%%%%%%%%%%%%%%%%%%%%%%%%%%%%%%%%%%%%%%%%%%%%%%%%%%% 
%%%%% ** Figure ** %%%%%%%%%%%%%%%%%%%%%%%%%%%%%%%%%%%%%%%%%%%%%%%%%%%
\begin{figure}[t]
	\begin{center}
	  \includegraphics[height=8cm]{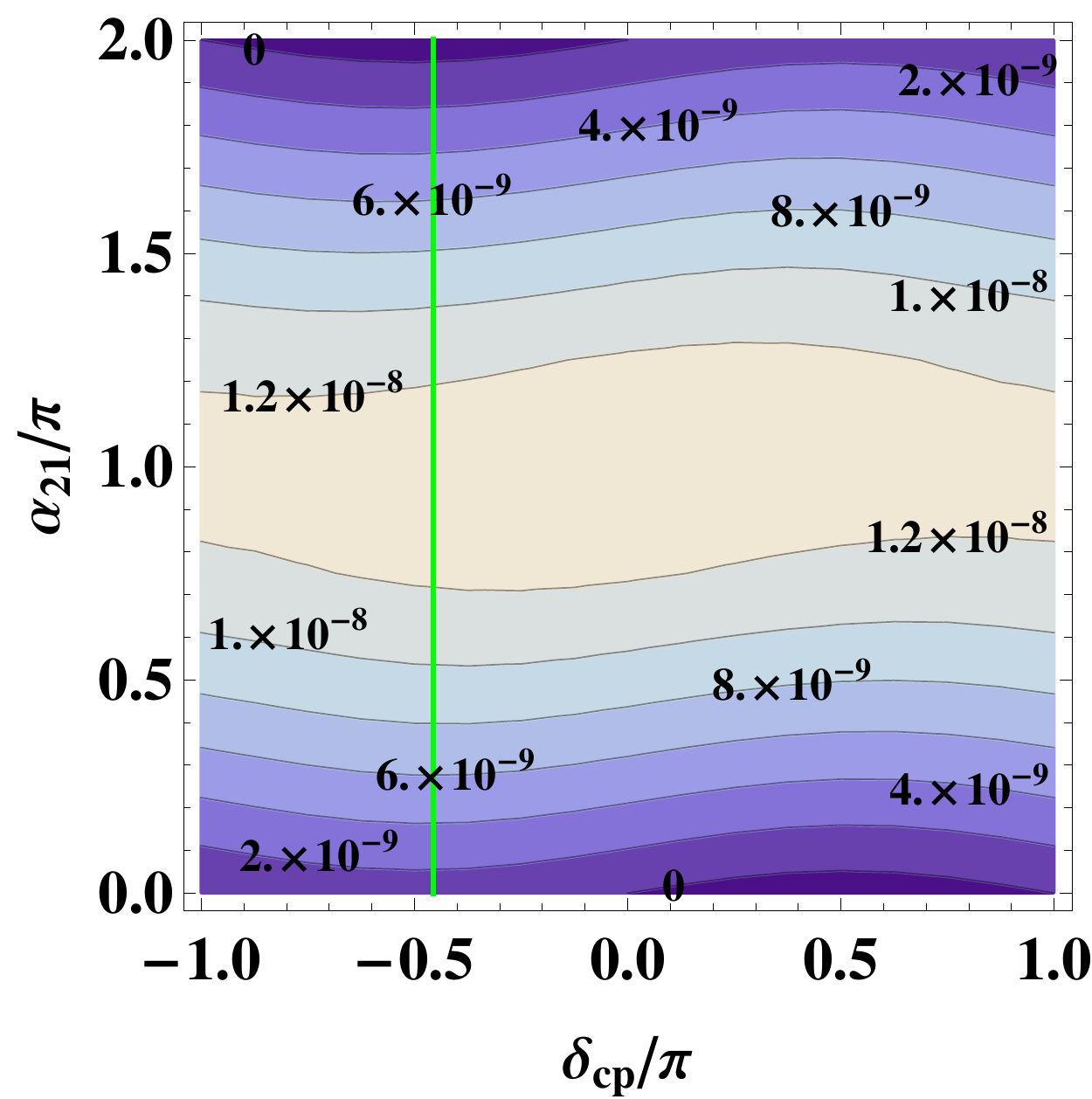}
	  \hspace{0.7cm}
    \includegraphics[height=8cm]{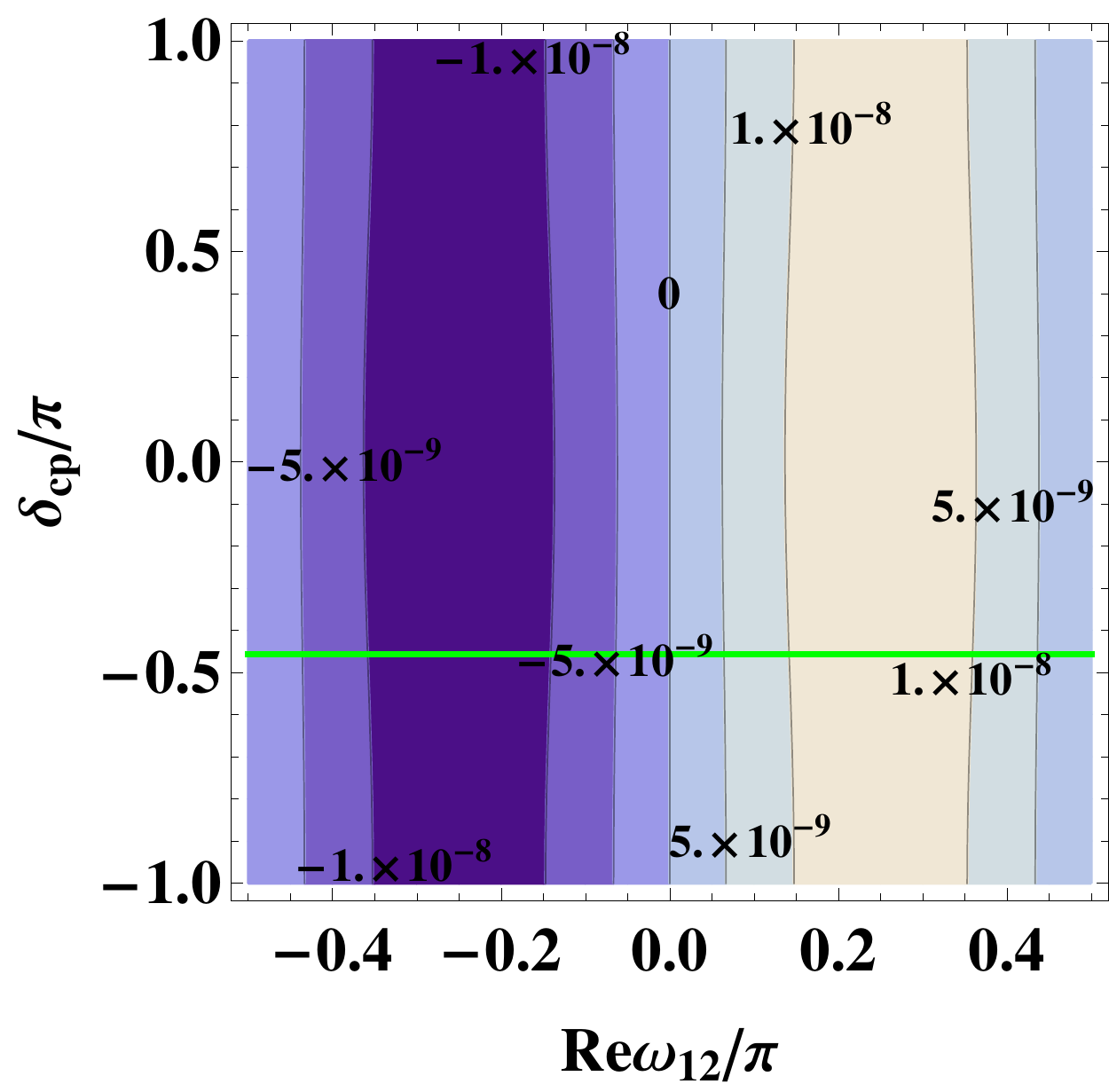}
  \caption{
	  Contour plots of $Y_B^{\rm MAX}$ in the IH case.
    $Y_B^{\rm MAX}$ is larger or smaller in the region with bright or 
    dark color.
    In the left or right panel the plot is shown in
    the $\delta_{\rm CP}$-$\alpha_{21}$ plane 
    when Re$\omega_{12} =  \pi/4$ or the Re$\omega_{12}$-$\delta_{\rm CP}$ 
    plane when $\alpha_{21} = \pi$, respectively.
		The green lines shows the central value of $\delta_{\rm CP}$
		from the global neutrino oscillation analysis.
  }
  \label{FIG_YB_IH}
  \end{center}
\end{figure}
%%%%%%%%%%%%%%%%%%%%%%%%%%%%%%%%%%%%%%%%%%%%%%%%%%%%%%%%%%%%%%%%%%%%%%
Next, we turn to consider the IH case.
It is found from Fig.~\ref{FIG_YB_IH} that $Y_B^{\rm MAX}$ is at most ${\cal O}(10^{-8})$, and 
hence resonant leptogenesis in the IH case is less effective compared with the NH case.
Moreover, the dependence on the CP phases are different from the NH case.
In the left panel of Fig.~\ref{FIG_YB_IH}
the contour plot of $Y_B^{\rm MAX}$ is shown in 
the $\delta_{\rm CP}$-$\alpha_{21}$ plane when $\mbox{Re}\omega_{21} = \pi/4$.
We find that $Y_B$ depends on the Majorana phase significantly as in the NH case, however the dependence on the Dirac phase is much milder than the NH case.
This behavior can also be seen in the right panel, 
which shows the contour plot of $Y_B^{\rm MAX}$ in the 
Re$\omega_{12}$-$\delta_{\rm CP}$ plane when $\alpha_{21} = \pi$.
It is found that the dependence on the CP phases are approximately given by
\begin{align}
Y_B \propto \sin \left( \frac{\alpha_{12}}{2} \right) \,.
\end{align}
Note that the subleading effect which disturbs the above dependence 
is larger than that in the NH case.
The observed BAU cannot be produced 
for the vanishing mixing between $\nu_R$'s at Re$\omega_{12}=0$, $\pi/2$
similar to the NH case.  In addition, the sign of the BAU correlates with the sign of Re$\omega_{12}$.

%%%%%%%%%%%%%%%%%%%%%%%%%%%%%%%%%%%%%%%%%%%%%%%%%%%%%%%%%%%%%%%%%%%%%% 
%%%%% ** Figure ** %%%%%%%%%%%%%%%%%%%%%%%%%%%%%%%%%%%%%%%%%%%%%%%%%%%
\begin{figure}[h]
	\begin{center}
    \includegraphics[height=8cm]{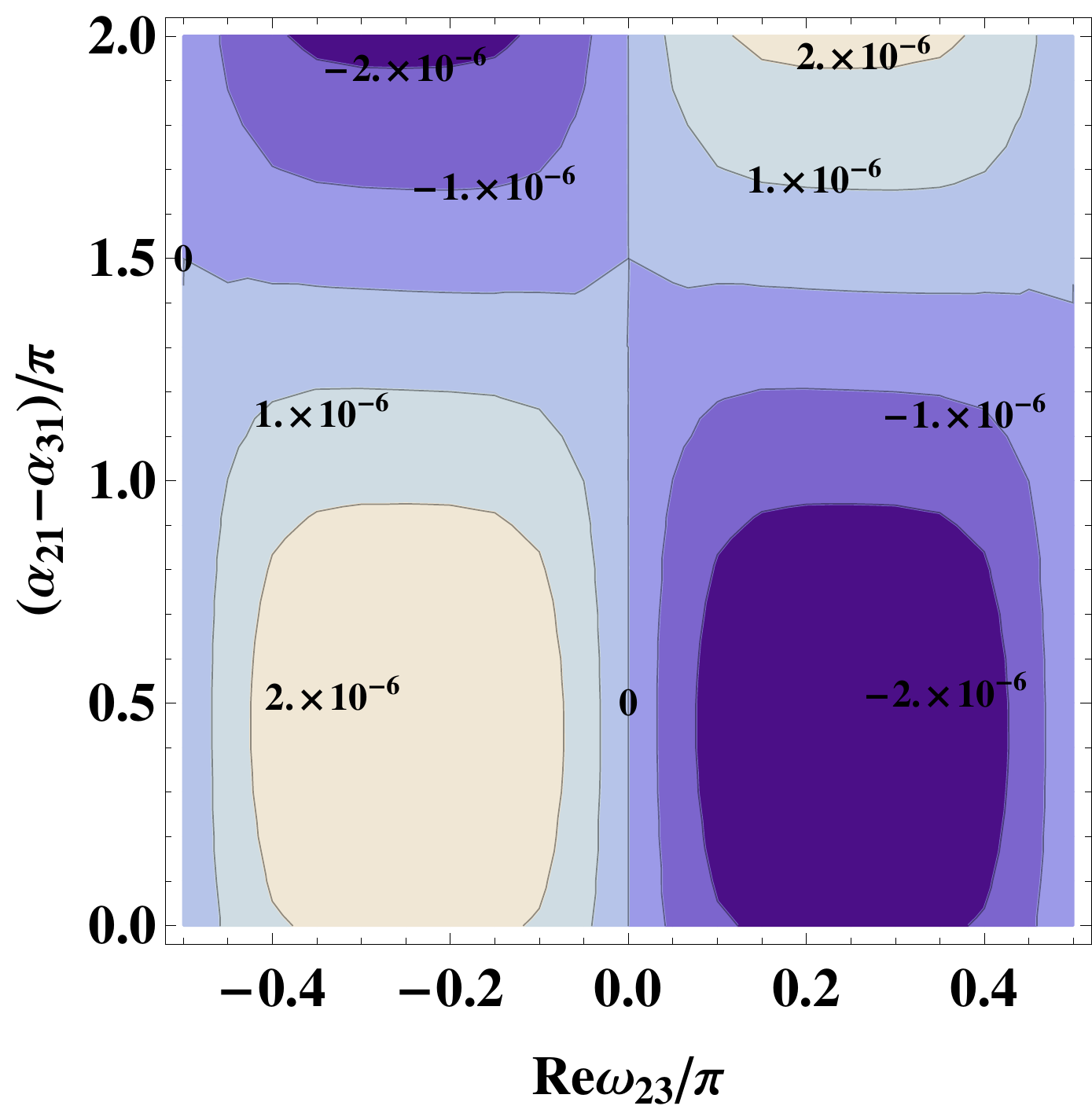}
	\hspace{0.7cm}
    \includegraphics[height=8cm]{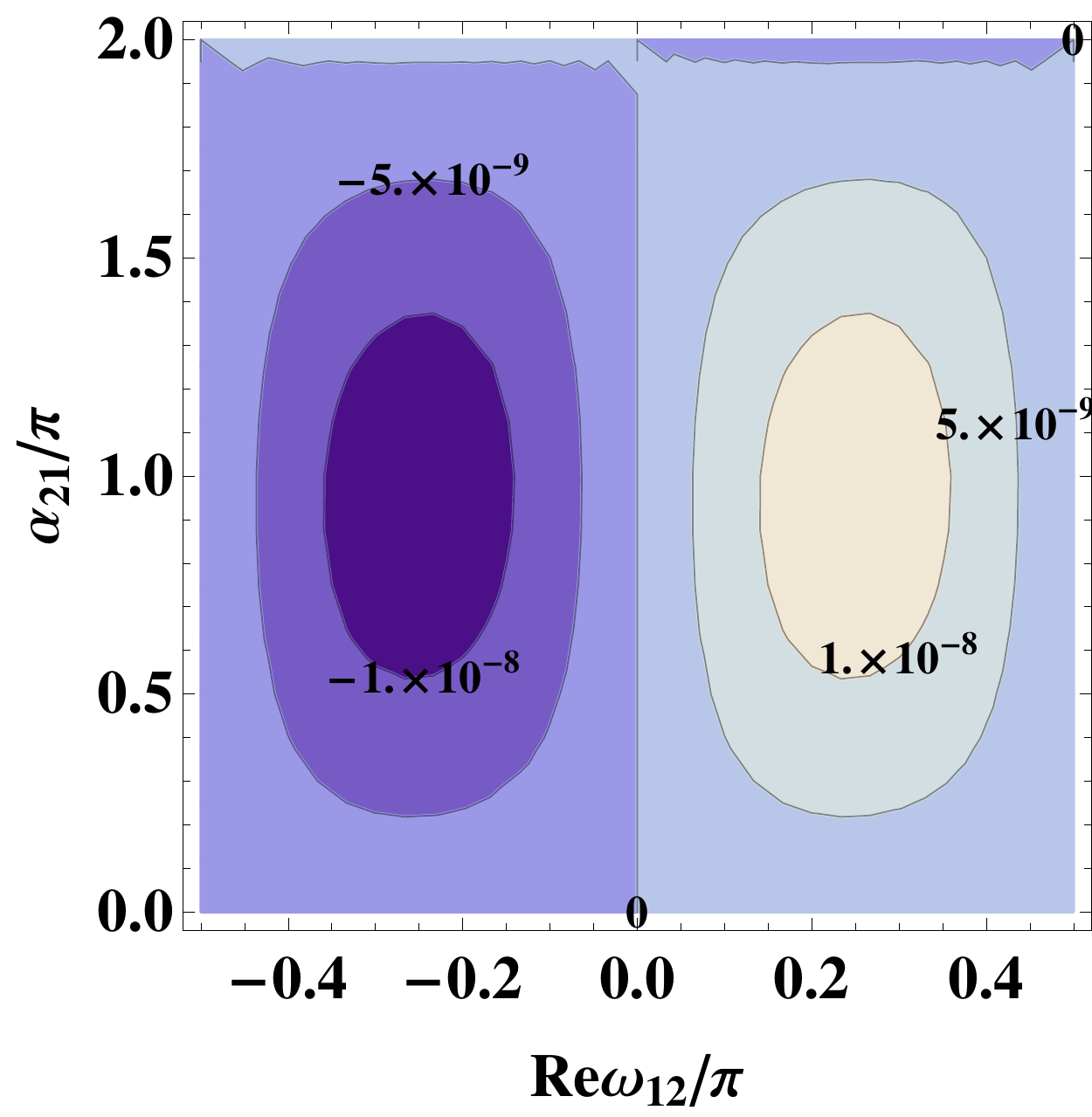}
  \caption{
    Contour plots of $Y_B^{\rm MAX}$ with 
    the Dirac phase in Eq.~(\ref{eq:DEL_cen}) 
    for the NH (left panel) and IH (right panel) cases.
    $Y_B$ is positive or negative in the 
    region with bright or dark color.
  }
  \label{FIG_YB_del}
  \end{center}
\end{figure}
%%%%%%%%%%%%%%%%%%%%%%%%%%%%%%%%%%%%%%%%%%%%%%%%%%%%%%%%%%%%%%%%%%%%%%
As described above, $Y_B^{\rm MAX}$ depends on the Dirac and Majorana
phases and the mixing angle of $\nu_R$'s.
We then discuss the case with the Dirac phase 
which is the central value from the global neutrino oscillation analysis
in Ref.~\cite{Esteban:2016qun}:
\begin{align}
	\label{eq:DEL_cen}
	\delta_{\rm CP} = 
	\left\{
		\begin{array}{l l}
			{}- 0.700 \, \pi~~~(234~^\circ) & ~~~\mbox{for the NH case}
			\\
			{}- 0.456 \, \pi~~~(278~^\circ) & ~~~\mbox{for the IH case}
		\end{array}
	\right. \,.
\end{align}
In this case the sign of the BAU is determined by
the Majorana phase and the mixing angle of $\nu_R$'s, which is represented 
in Fig.~\ref{FIG_YB_del}.
The left panel shows the contour plot of $Y_B^{\rm MAX}$ in the 
$\mbox{Re}\omega_{23}$-$(\alpha_{21}-\alpha_{31})$ plane for the NH case.
Interestingly, we observe that 
the correct value of $Y_B$ can be realized for all possible values
of Majorana phase by choosing the appropriate angle $\mbox{Re}\omega_{23}$.
On the other hand, for the IH case
the contour plot of $Y_B^{\rm MAX}$ in the
$\mbox{Re}\omega_{12}$-$\alpha_{21}$ plane is shown in 
the right panel of Fig.~\ref{FIG_YB_del}.
It is seen that the successful baryogenesis is realized for both $\mathrm{Re}\omega_{12}<0$ and  $\mathrm{Re}\omega_{12}>0$. When $\mathrm{Re}\omega_{12}<0$, $\alpha_{21} \simeq 2\pi$ is required and then the CP violation by $\delta_{\rm CP}$ is essential. 

As explained above, the Majorana phase plays an important
role for determining the sign of the BAU through leptogenesis scenario 
under consideration.  It is therefore expected that 
the BAU may give an impact on the other phenomena in which the Majorana phase
is essential.  One such example is the $0 \nu \beta \beta$ decay, which will be discussed in the next section.

Before closing this section, we should mention the dependence on 
the averaged Majorana mass $M_N$ of $\nu_R$'s.
It is interesting to note that the CP asymmetry parameter 
with $\Delta M = \Delta M_\ast$
does not depend on $M_N$ in the considering situation.
This results in the fact that $Y_B^{\rm MAX}$ is almost insensitive to $M_N$ as long as $M_N \lesssim 30$~TeV.
In such a mass region the small dependence on $M_N$ arises from the relative size
between $M_N$ and the sphaleron's freeze-out temperature $T_{\rm sph}$.
When $M_N$ in the TeV region, leptogenesis occurs at
$T \sim M_N$ which is not far from the $T_{\rm sph}$.
This means that the conversion of the lepton asymmetry into the baryon 
asymmetry terminates in the course of the washout processes.
Thus, the final BAU does depend on $M_N$.  If $M_N$ is sufficiently 
larger than $T_{\rm sph}$, this effect vanishes because $Y_B$ is frozen well before $T = T_{\rm sph}$.
In addition, $Y_B^{\rm MAX}$ suffers from the effects
of the scattering processes, which induces a small dependence of $M_N$.
On the other hand, when $M_N \gtrsim 30$~TeV, 
our assumption that the processes by the Yukawa interactions for all quarks and leptons including electrons are in thermal equilibrium at the leptogenesis regime is broken \cite{Garbrecht:2014kda}.
In such cases, the treatment of the flavor effects must be changed, which leads to a considerable modification of the estimation of $Y_B^{\rm MAX}$.
Therefore, our results in the present analysis are also insensitive
to the choice of $M_N$ as long as $M_N$ is sufficiently small.
%%%%%%%%%%%%%%%%%%%%%%%%%%%%%%%%%%%%%%%%%%%%%%%%%%%%%%%%%%%%%%%%%%%%%%
\section{Neutrinoless double beta decay}
%%%%%%%%%%%%%%%%%%%%%%%%%%%%%%%%%%%%%%%%%%%%%%%%%%%%%%%%%%%%%%%%%%%%%%
In the seesaw mechanism active neutrinos are Majorana fermions, and
the lepton number is broken in contrast to the SM.
One interesting example of the lepton number violating processes
is the $0\nu \beta \beta$ decay: 
$(A,Z) \to (A,Z+2) + 2 e^-$~\cite{Pas:2015eia}.
The decay rate is proportional to $m_{\rm eff}^2$ where the effective neutrino mass given by
\begin{align}
  m_{\rm eff} = \bigg| \sum_{i} m_i U_{ei}^2 \bigg| \,.
\end{align}
Here we take into account the contribution only from active neutrinos,
because that from heavy neutral leptons
is negligible in the considered situation.

%%%%%%%%%%%%%%%%%%%%%%%%%%%%%%%%%%%%%%%%%%%%%%%%%%%%%%%%%%%%%%%%%%%%%%
%%%%% ** Figure ** %%%%%%%%%%%%%%%%%%%%%%%%%%%%%%%%%%%%%%%%%%%%%%%%%%%
\begin{figure}[t]
  \centerline{
    \includegraphics[height=8cm]{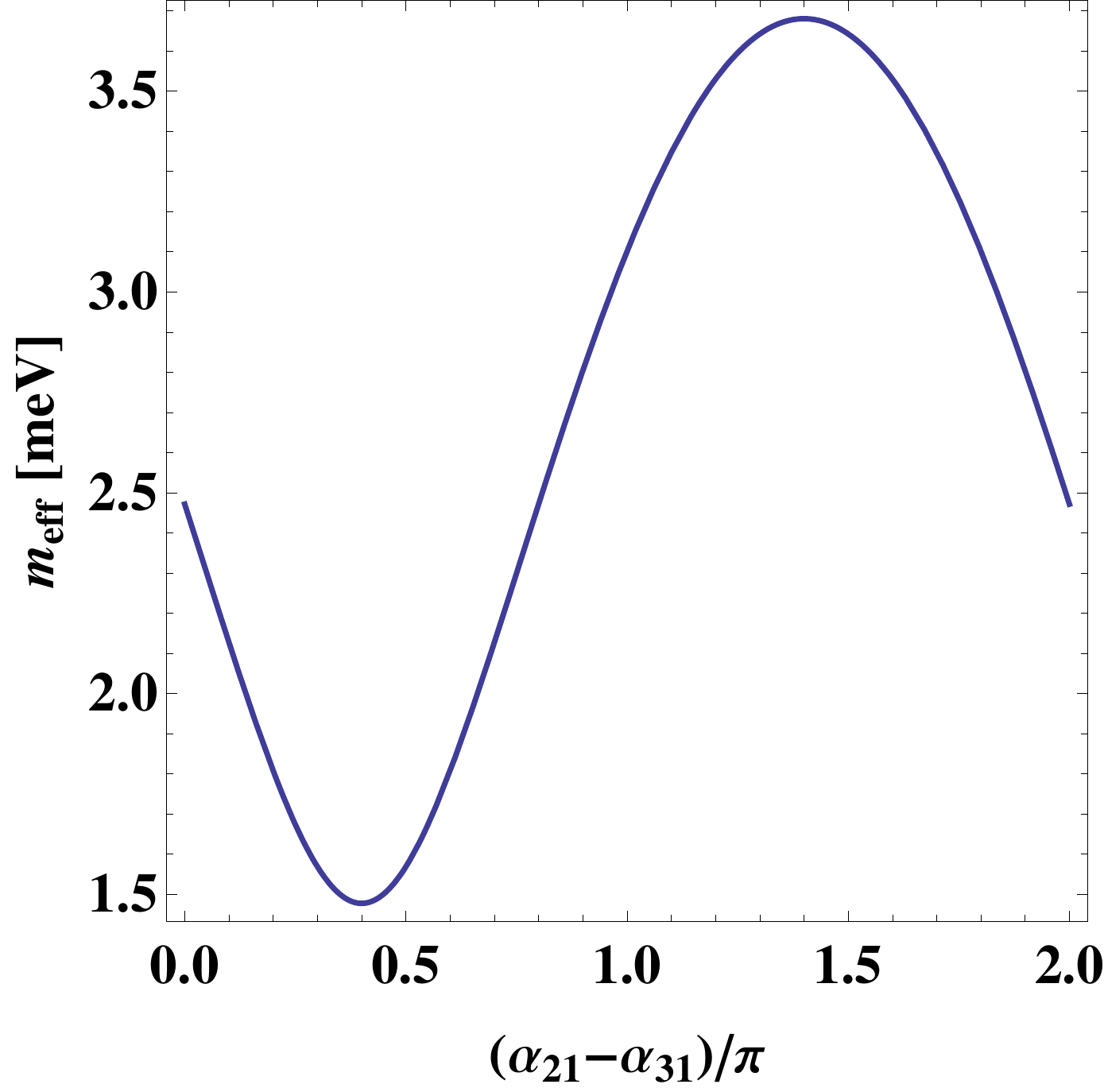}	
	\hspace{0.7cm}
    \includegraphics[height=8cm]{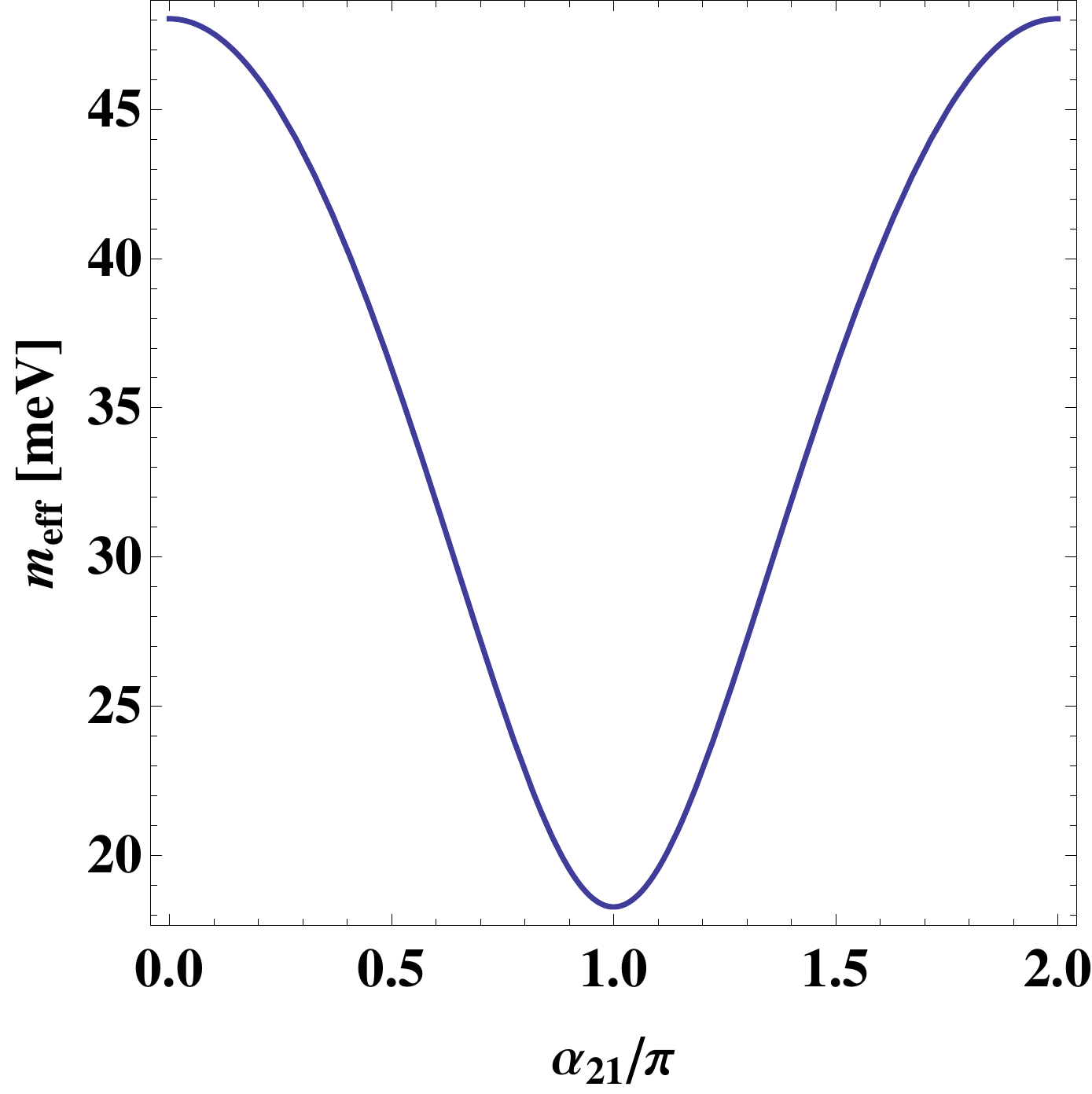}
  }%
  \caption{
    Effective neutrino mass $m_{\rm eff}$ of neutrinoless double beta decay
    in terms of the Majorana phase $(\alpha_{21}-\alpha_{31})$ for the NH case
    (left panel) and $\alpha_{21}$ for the IH case (right panel),
    respectively.
  }
  \label{FIG_meff}
\end{figure}
%%%%%%%%%%%%%%%%%%%%%%%%%%%%%%%%%%%%%%%%%%%%%%%%%%%%%%%%%%%%%%%%%%%%%%
We have assumed that only two right-handed neutrinos are responsible
to the seesaw mechanism of neutrino masses, and then 
the lightest active neutrino is massless.
In this case the effective mass is written as
\begin{align}
  m_{\rm eff}^2 =
  m_2^2 \, c_{13}^4 \, s_{12}^4 + m_3^2 \, s_{13}^4 +
  2 \, m_2 \, m_3 \, c_{13}^2 \, s_{12}^2 \, s_{13}^2
  \cos(\alpha_{21}-\alpha_{31}+2\delta_{\rm CP}) \,,
\end{align}
for the NH case and
\begin{align}
  m_{\rm eff}^2 = 
  c_{13}^4 \,
  \left[ m_1^2 \, c_{12}^4 +
    m_2^2 \, s_{12}^4 + 2 \, m_1 \, m_2 \, c_{12}^2 \, s_{12}^2 \,
    \cos\alpha_{21}
  \right]
 \,,
\end{align}
for the IH case.
It is seen that $m_{\rm eff}$ depends on mixing angles and masses of active 
neutrinos as well as the CP violating phases.
The effective mass when we use the central value of $\delta_{\rm CP}$
in Eq.~(\ref{eq:DEL_cen})
can be determined by the Majorana phase as shown in Fig.~\ref{FIG_meff}.
It is found that the possible range is 
\begin{align}
	m_{\rm eff} =
	\left\{
		\begin{array}{l l}
			(1.5-3.7)~\mbox{meV}
		  &~~~\mbox{for the NH case}
			\\
			(18-48)~\mbox{meV}
		  &~~~\mbox{for the IH case}
		\end{array}
	\right. \,.
\end{align}
where the minimal and maximal values of $m_{\rm eff}$ are achieved 
if  $\alpha_{21}-\alpha_{31} = 0.4 \, \pi$ and $1.4 \, \pi$ for the NH case,
and
$\alpha_{21} = \pi$ and $0$ for the IH case, respectively.

We have shown in the previous section
that the successful scenario for resonant leptogenesis 
requires a certain range of the Majorana phase 
(\ie, $\alpha_{21}-\alpha_{31}$ or $\alpha_{21}$ for the NH or IH case)
as well as the mixing angles of $\nu_R$'s
(\ie, Re$\omega_{23}$ or Re$\omega_{12}$ 
for the NH or IH case).  It is, therefore, expected that 
the predicted range of $m_{\rm eff}$ is restricted for generating
the sufficient amount of the BAU by leptogenesis.

%%%%%%%%%%%%%%%%%%%%%%%%%%%%%%%%%%%%%%%%%%%%%%%%%%%%%%%%%%%%%%%%%%%%%%
%%%%% ** Figure ** %%%%%%%%%%%%%%%%%%%%%%%%%%%%%%%%%%%%%%%%%%%%%%%%%%%
\begin{figure}[t]
	\begin{center}
    \includegraphics[height=8cm]{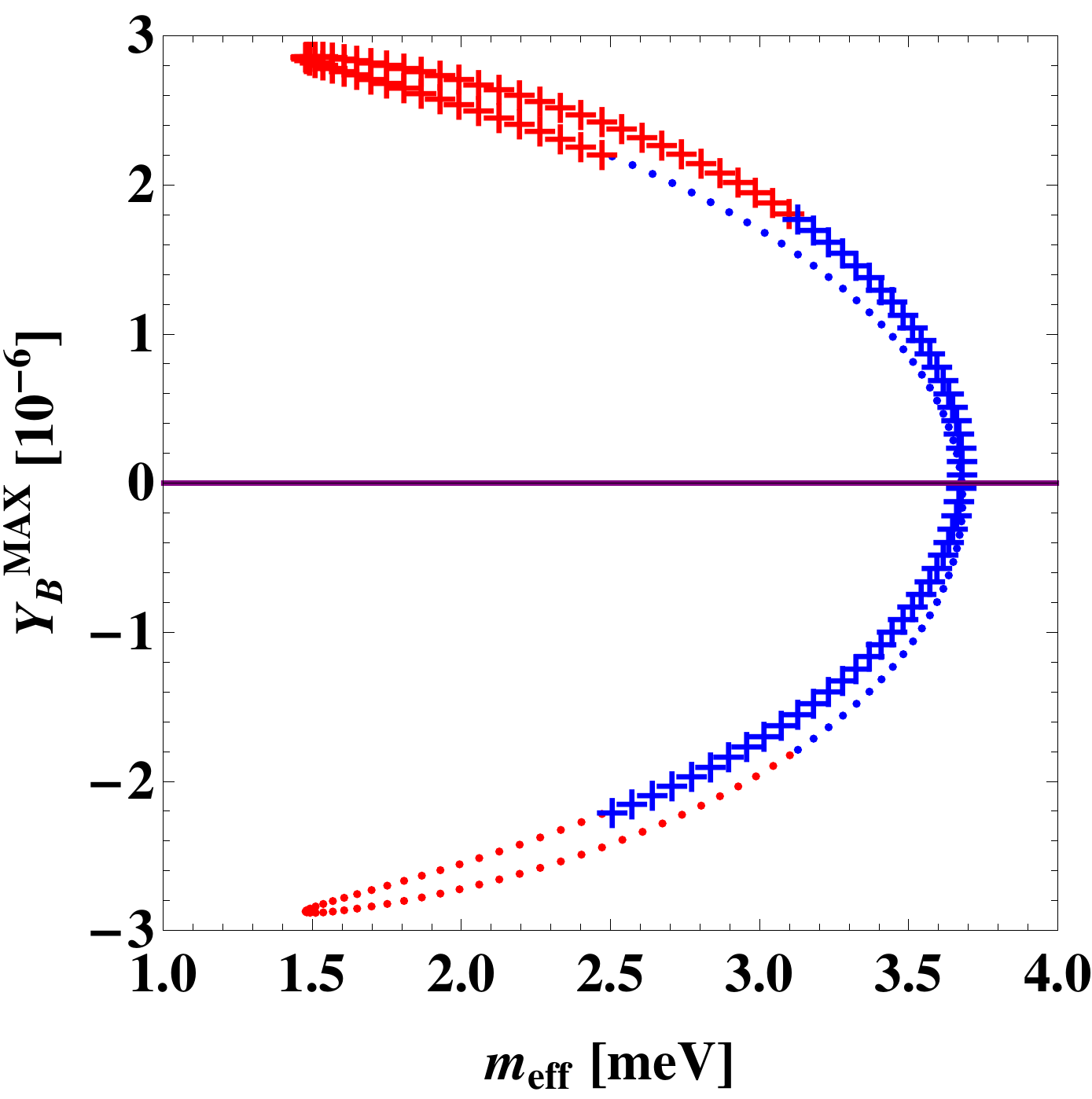}
	\hspace{0.7cm}
    \includegraphics[height=8cm]{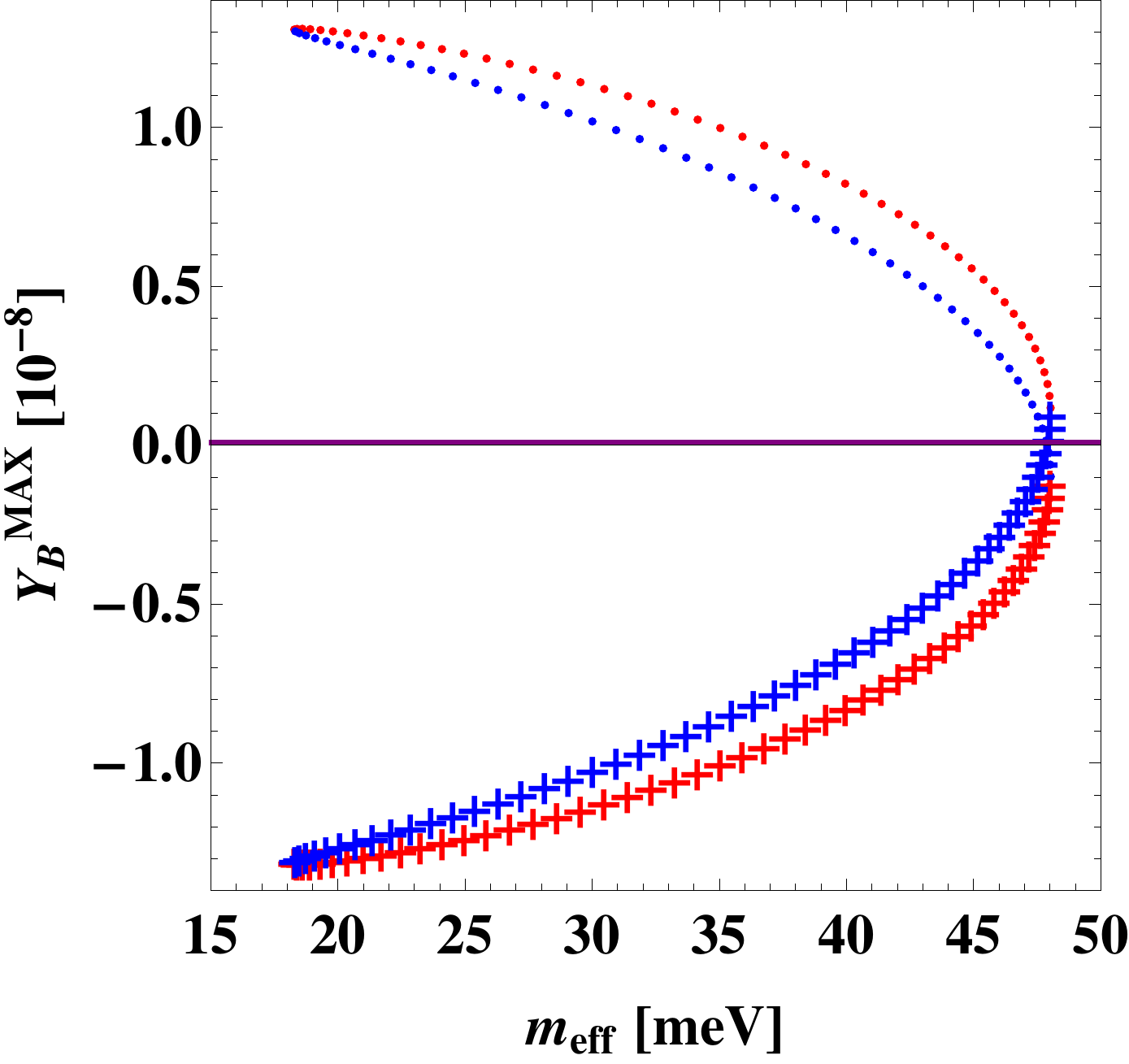}
  \caption{
    Yield of the BAU $Y_B^{\rm MAX}$ in terms of effective neutrino mass $m_{\rm eff}$.
    In the left panel we consider the NH case
    and take
    Re$\omega_{23} = - \pi/4$ and $(\alpha_{21}-\alpha_{31})= 0$ to $\pi$ (red cross marks),
    Re$\omega_{23} = + \pi/4$ and $(\alpha_{21}-\alpha_{31})= 0$ to $\pi$ (red dot marks),
    Re$\omega_{23} = - \pi/4$ and $(\alpha_{21}-\alpha_{31})= \pi$ to $2 \pi$ (blue cross mark), and
    Re$\omega_{23} = + \pi/4$ and $(\alpha_{21}-\alpha_{31})= \pi$ to $2 \pi$ (blue dot marks).
    In the right panel we consider the IH case
    and take
    Re$\omega_{12}=-\pi/4$ and $\alpha_{21}=0$ to $\pi$ (red cross marks),
    Re$\omega_{12}=+\pi/4$ and $\alpha_{21}=0$ to $\pi$ (red dot marks),
    Re$\omega_{12}=-\pi/4$ and $\alpha_{21}=\pi$ to $2 \pi$ (blue cross mark), and
    Re$\omega_{12}=+\pi/4$ and $\alpha_{21}=\pi$ to $2 \pi$ (blue dot marks).
    The horizontal lines are the observed value of $Y_B^{\rm OBS}$.
  }
  \label{FIG_YB_meff}
  \end{center}
\end{figure}
%%%%%%%%%%%%%%%%%%%%%%%%%%%%%%%%%%%%%%%%%%%%%%%%%%%%%%%%%%%%%%%%%%%%%% 
It is important to note that $Y_B^{\rm MAX}$ and $m_{\rm eff}$ depend on 
an unique unknown parameter, \ie, the Majorana phase, for a given the mixing 
angle Re$\omega_{23}$ or Re$\omega_{12}$,
and hence we can obtain the nontrivial relation between these parameters.
This point is illustrated in Fig.~\ref{FIG_YB_meff}.
Namely, by changing the value of the mixing angle,
a locus can be described in the $m_{\rm eff}$-$Y_B^{\rm MAX}$ plane.

\begin{figure}[t]
	\begin{center}
    \includegraphics[height=8cm]{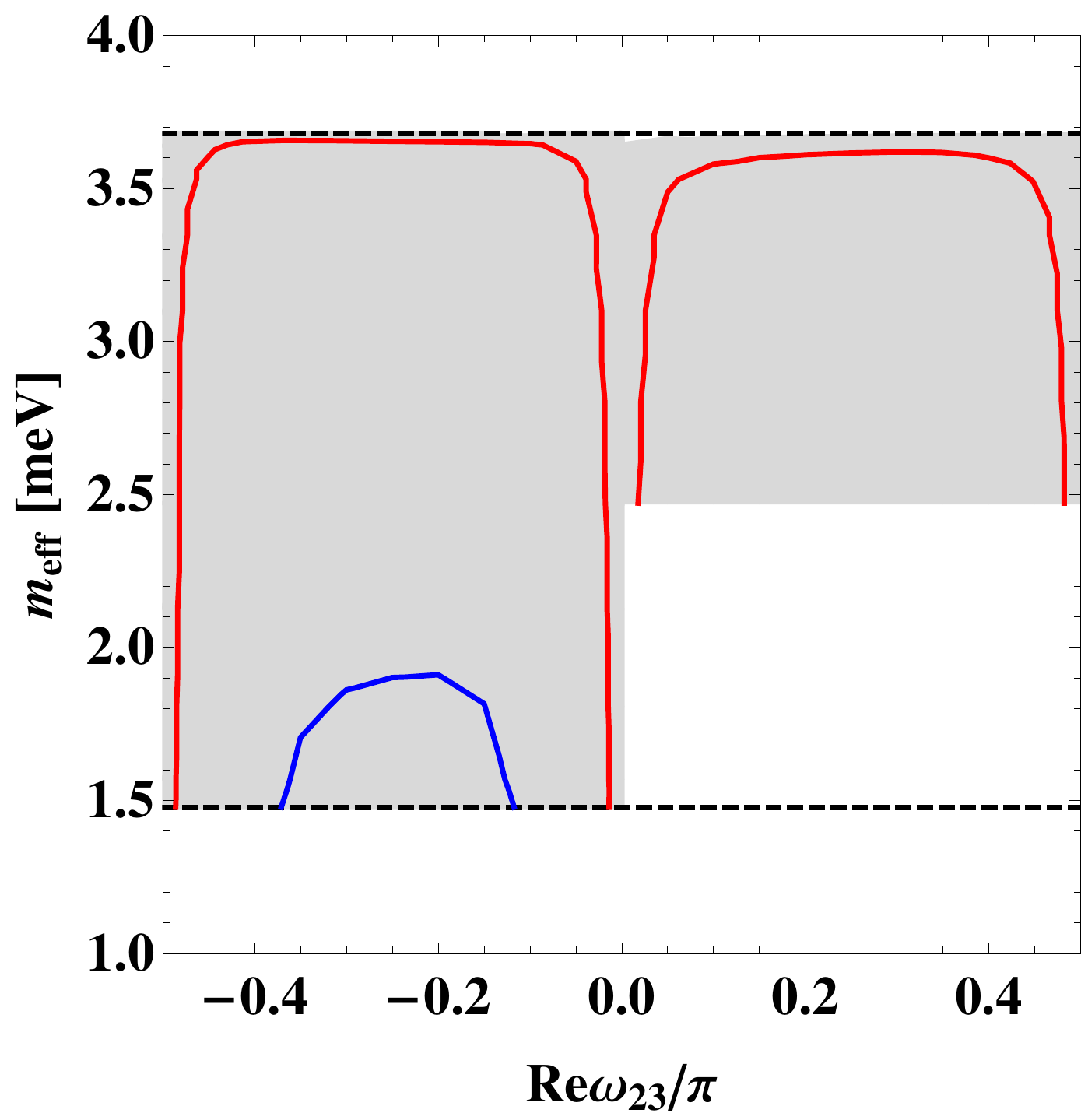}
	\hspace{0.7cm}
	  \includegraphics[height=8cm]{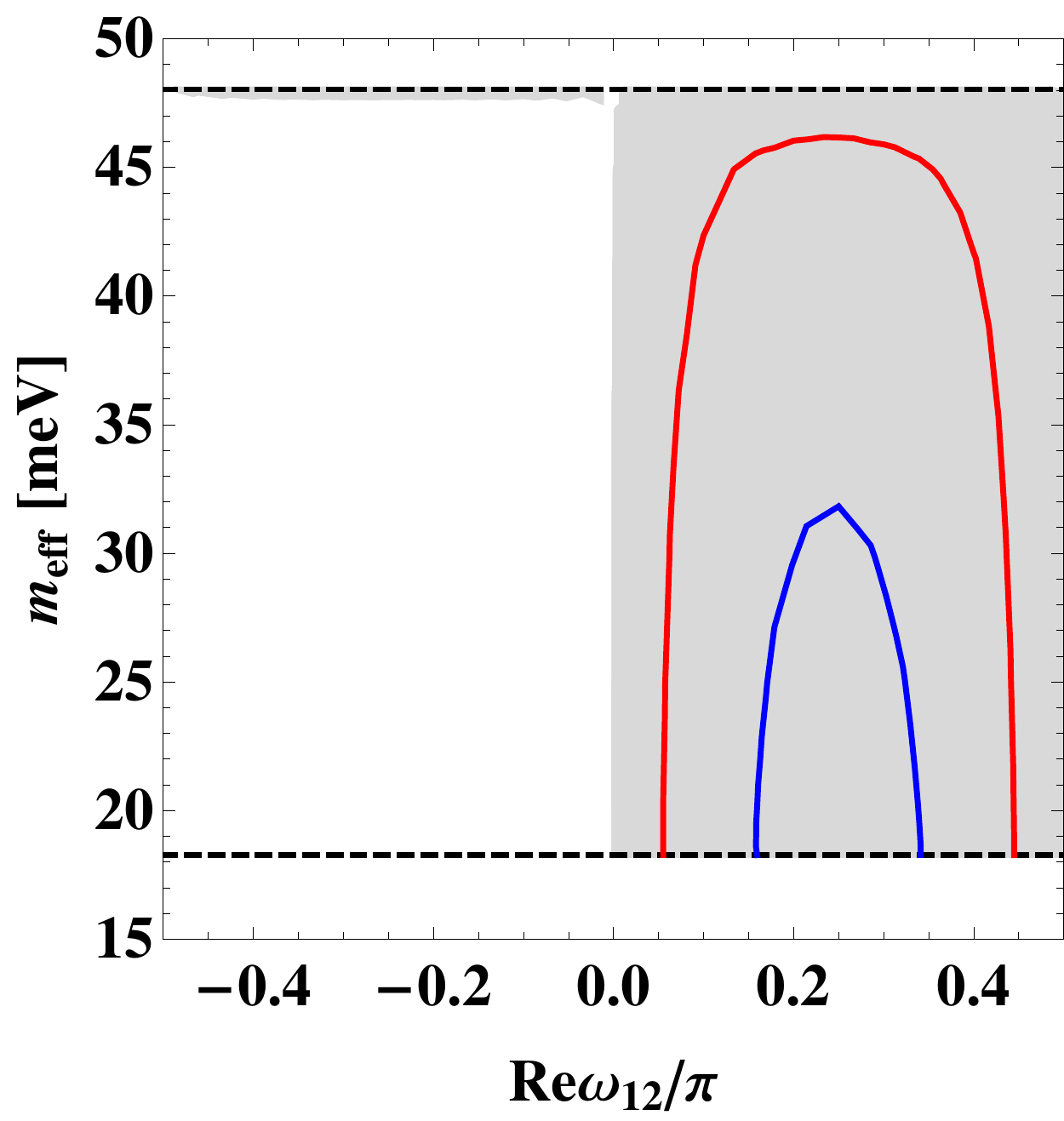}
  \caption{
   Range of effective neutrino mass $m_{\rm eff}$ in order
    to account for the observed BAU 
	for the NH (left panel) and IH (right panel) cases.
	The colored region is allowed.
    In the NH case,  the upper bounds for $\Delta M = 1.0 \times 10^4 \Delta M_\ast$ and $\Delta M = 6.0 \times 10^4 \Delta M_\ast$ are shown by red and blue lines, respectively. In the IH case, the upper bounds for $\Delta M = 1.0 \times 10^2 \Delta M_\ast$ and $\Delta M = 2.5 \times 10^2 \Delta M_\ast$ are shown by red and blue lines, respectively.
  }
  \label{FIG_meff_rom}
 	\end{center}
\end{figure}

In Fig.~\ref{FIG_meff_rom} we show the predicted region of $m_{\rm eff}$
in terms of the mixing angle in order to account for the observed BAU.
For the NH case the impact on the $0\nu \beta \beta$ decay is different 
depending on values of Re$\omega_{23}$.  
When Re$\omega_{23} <0$, the prediction of $m_{\rm eff}$ is unaffected by 
the BAU.  On the other hand, when Re$\omega_{23} >0$, 
$m_{\rm eff}$ receives the lower bound from the BAU.
Thus, although the absolute upper and lower bounds are not changed by the BAU
without knowing Re$\omega_{23}$,
the range of $m_{\rm eff}$ may become smaller substantially if the positive value of Re$\omega_{23}$ would be realized. Note that the BAU gives the upper bound on $m_{\rm{eff}}$ when $\Delta M$ becomes sufficiently large, and the allowed region disappears for $\Delta M \geq \mathcal{O} (10^5) \Delta M_{\ast}$

On the other hand, the successful baryogenesis is realized for $\rm{Re} \omega_{12} \ne 0$ and $\pi/2$ for the IH case. When $\rm{Re}\omega_{12}<0$, the BAU suggests $\alpha_{21} \simeq 2\pi$, which leads to the maximal value of $m_{\rm {eff}}$. 
Moreover, as mentioned above, 
$m_{\rm eff}$ receives an additional upper bound from the BAU.
As shown in Fig.~\ref{FIG_meff_rom}, the possible region vanishes for 
$\Delta M \gtrsim 300 \, \Delta M_\ast$.

We have so far taken the Dirac phase in Eq.~(\ref{eq:DEL_cen}).
The allowed region in Fig.~6 changes by the value of $\delta_{\rm CP}$ for the NH case, while it remains almost the same for the IH case. 
When Re$\omega_{23}>0$, the lower bound on the effective mass becomes severer as $\delta_{\rm CP}$ becomes close to 200$^\circ$, while weaker as $\delta_{\rm CP}$ becomes close to 270$^\circ$. 
On the other hand, when Re$\omega_{23} <0$, the allowed range of the effective mass remains unchanged in the $1 \sigma$ range ($\delta_{\rm CP} = 203^\circ-~277^\circ$~\cite{Esteban:2016qun}).
This difference comes from the $\delta_{\rm CP}$ dependence of $Y_B$. 
The experimental determination of $\delta_{\rm CP}$ is thus important for the predictions of the BAU as well as the $0 \nu \beta \beta$ decay for the NH case.
%%%%%%%%%%%%%%%%%%%%%%%%%%%%%%%%%%%%%%%%%%%%%%%%%%%%%%%%%%%%%%%%%%%%%%
%%%%% ** Figure ** %%%%%%%%%%%%%%%%%%%%%%%%%%%%%%%%%%%%%%%%%%%%%%%%%%%
%%%%%%%%%%%%%%%%%%%%%%%%%%%%%%%%%%%%%%%%%%%%%%%%%%%%%%%%%%%%%%%%%%%%%%

%%%%%%%%%%%%%%%%%%%%%%%%%%%%%%%%%%%%%%%%%%%%%%%%%%%%%%%%%%%%%%%%%%%%%%
\section{Conclusions}
%%%%%%%%%%%%%%%%%%%%%%%%%%%%%%%%%%%%%%%%%%%%%%%%%%%%%%%%%%%%%%%%%%%%%%
We have investigated the Standard Model extended by right-handed neutrinos,
in which two $\nu_R$'s are quasi-degenerate with TeV-scale masses.
We have studied the origins of the neutrino masses and the BAU in this setup.
By assuming that the neutrino masses are generated by the seesaw mechanism,
the production of the BAU through resonant leptogenesis has been 
studied by using the Boltzmann equations with the flavor effects.

We have considered the case when the CP violation occurs only in the 
active neutrino sector.  Since leptogenesis by TeV-scale right-handed neutrinos
is considered, the flavor effects are essential to obtain the non-zero BAU.
For the two right-handed neutrino case, the parameters
relevant for the sign of the BAU are 
the Dirac phase $\delta_{\rm CP}$, the Majorana phase 
$\alpha_{21}-\alpha_{31}$ or $\alpha_{21}$ for the NH or IH case,
and the mixing parameter Re$\omega_{23}$ or Re$\omega_{12}$ 
for the NH or IH case.  We have shown how the BAU 
depends on these parameters and identified the possible range
of these parameters to account for the observed BAU.

Furthermore, we have discussed the impact on the $0 \nu \beta \beta$ decay.
It has been found that the predicted range of the effective neutrino mass
$m_{\rm eff}$ depends significantly on the mixing parameter and mass difference of $\nu_{R}$'s as well as
the mass hierarchy of active neutrinos. Especially, when $\Delta M$ becomes sufficiently larger than $\Delta M_{\ast}$, $m_{\rm {eff}}$ receives the stringent upper bound in order to account for the BAU.

%%%%%%%%%%%%%%%%%%%%%%%%%%%%%%%%%%%%%%%%%%%%%%%%%%%%%%%%%%%%%%%%%%%%%% 
\section*{Acknowledgments}
%%%%%%%%%%%%%%%%%%%%%%%%%%%%%%%%%%%%%%%%%%%%%%%%%%%%%%%%%%%%%%%%%%%%%%
The work of TA was partially supported by JSPS KAKENHI Grant Numbers
17H05198, 17K05410, and 18H03708.
TA and TY thank the Yukawa Institute for
Theoretical Physics at Kyoto University for the useful discussions
during "the 23th Niigata-Yamagata joint school" (YITP-S-18-03).

\appendix
%%%%%%%%%%%%%%%%%%%%%%%%%%%%%%%%%%%%%%%%%%%%%%%%%%%%%%%%%%%%%%%%%%%%%%
\section{Boltzmann equations}	\label{boltzmanneq}
%%%%%%%%%%%%%%%%%%%%%%%%%%%%%%%%%%%%%%%%%%%%%%%%%%%%%%%%%%%%%%%%%%%%%%
In this appendix we present the formulae to estimate the yield of the BAU
by using the Boltzmann equations.
The yield of the right-handed neutrino $N_I$, $Y_{N_I}$, is given by
\begin{align}
  Y_{N_I} = \frac{n_{N_I}}{s} \,,
\end{align}
where $n_{N_I}$ is the number density of $N_I$.
The entropy density of the universe is 
$s = (2 \pi^2/45) g_{\ast s} T^3$ with the cosmic temperature $T$.
The effective degrees of freedom is taken as
$g_{\ast s} = 106.75$ throughout this analysis.
When $N_I$ is in equilibrium, the yield is estimated as
\begin{align}
  Y_{N_I}^{\rm eq} = \frac { 45 } { 2 \pi ^ { 4 } g _ { * s } } (M_I^2/M^2) z ^ { 2 } K _ { 2 } \left( (M_I/M) z \right) \,,
\end{align}
where the variable $z$ is defined by $z = M/T$, in which 
$M$ is the mass of lighter right-handed neutrinos, \ie,
$M = M_2$ or $M_1$ for the NH or IH case, respectively.
$K_2(z)$ is the modified Bessel function of the second kind.
When the sphaleron processes are in thermal equilibrium,
there are three conserved charges
$X_\alpha = B/3 - L_\alpha$ ($\alpha = e, \mu, \tau$)
associated with baryon number $B$ and lepton flavors $L_{\alpha}$.
The yield of $X_\alpha$ is denoted by $Y_{X_\alpha} = n_{X_\alpha}/s$.
The Boltzmann equations for $Y_{N_I}$ and $Y_{X_\alpha}$
used in this analysis are given by~\cite{Plumacher:1996kc,Plumacher:1998ex}
\begin{align}
  \frac{d Y_{N_I}}{dz}
  &=
    -\frac{z}{s H(M)}
    \left	\{	\left( \frac{Y_{N_I}}{Y_{N_I}^{\rm eq}}	-	1	\right)
    (\gamma_{N_I}^{} +2 \gamma_{N_I t}^{(1)}	+4\gamma_{N_I t}^{(2)})
    \ +	\sum_{J \neq I}
    \left(	\frac{Y_{N_I}}{Y_{N_I}^{\rm eq}}\frac{Y_{N_J}}{Y_{N_J}^{\rm eq}}	-	1	\right)
    (\gamma_{N_I N_J}^{(1)}+\gamma_{N_I N_J}^{(2)})
    \right \} \,,
    \label{eq:yieldN1}\\
  \frac{d Y_{X_{\alpha}}}{dz}
  &=
    -\frac{z}{s H(M)}	\left	\{	\sum_{I}
    \left(	\frac{Y_{N_I}}{Y_{N_I}^{\rm eq}}	-1\right)
    \varepsilon_{\alpha I} \gamma_{N_I}^{}
    -
    \sum_{\beta} \left[ \sum_{I}
    \left(	\frac{1}{2} \left(C_{\alpha \beta}^{\ell}	-	C_{\beta}^{\Phi}\right)
    \gamma_{N_1}^{\alpha}	\right.\right.\right.
    \notag	\\	
  &\left. \left.
    + \left(C_{\alpha \beta}^{\ell}\frac{Y_{N_I}}{Y_{N_I}^{\rm eq}}
    -	\frac{C_{\beta}^{\Phi}}{2}\right)\gamma_{N_1 t}^{(1)}
    + \left(	2C_{\alpha \beta}^{\ell}	-\frac{C_{\beta}^{\Phi}}{2}
    \left(	1+\frac{Y_{N_I}}{Y_{N_I}^{\rm eq}} \right) \right) \gamma_{N_1 t}^{(2)}
    \right) \right.
    \notag\\	
  &\left.\left.
    + \sum_{\gamma} \left(
    \left(	C_{\alpha \beta}^{\ell}	+C_{\gamma \beta}^{\ell}		-2C_{\beta}^{\Phi}\right)
    \left(	\gamma_{N}^{(1)\alpha \gamma}	+\gamma_{N}^{(2)\alpha \gamma}\right)
    +	\sum_{I,J}	\left(	C_{\alpha \beta}^{\ell}	-C_{\gamma \beta}^{\ell}\right)
    \gamma_{N_{I} N_{J}}^{(1) \alpha \gamma}	\right)
    \right] \frac{Y_{X_{\beta}}}{Y^{\rm eq}}
    \right \} \,,		\label{eq:Boltz_falavor}
\end{align}
where $H(M)$ is the Hubble expansion rate for
$T=M$ and $Y^{\rm eq}= 45/(2 \pi^4 g_{\ast s})$.%
\footnote{
We have applied the Bolzmann approximation.}
$C^{\ell}$ and $C^{\Phi}$ are given by~\cite{Nardi:2006fx,Abada:2006ea}
\begin{align}
  C^{\ell}&=\frac{1}{711}
  \begin{pmatrix}
    -211&	16&	16\\
    16&		-211&	16\\
    16&		16&		-211
  \end{pmatrix}\,,~~~~~
  C^{\Phi}=\frac{8}{79}
  \begin{pmatrix}
    1,&	1,&	1
  \end{pmatrix} \,.
\end{align}
The reaction densities $\gamma$'s are found in 
Refs.~\cite{Plumacher:1996kc,Plumacher:1998ex}.
We have considered the non-supersymmetric case and taken into account
the flavor effects of leptogenesis.  
Our notations of the reaction densities 
correspond to those in Refs.~\cite{Plumacher:1996kc,Plumacher:1998ex} as
$\gamma_{N}^{(1)} = \gamma_N^{(2)}$,
$\gamma_N^{(2)} = \gamma_N^{(13)}$,
$\gamma_{N_I t}^{(1)} = \gamma_{t I}^{(3)}$,
$\gamma_{N_I t}^{(2)} = \gamma_{t I}^{(4)}$,
$\gamma_{N_I N_J}^{(1)}= \gamma_{N_I N_J}^{(2)}$,
and $\gamma_{N_I N_J}^{(2)} = \gamma_{N_I N_J}^{(3)}$.
The yield of the BAU is then given by 
\begin{align}
Y_{B}&=\frac{28}{79} \left. \sum_{\alpha} Y_{X_{\alpha}}\right|_{T=T_{\rm sph}}.
\end{align}

%%%%%%%%%%%%%%%%%%%%%%%%%%%%%%%%%%%%%%%%%%%%%%%%%%%%%%%%%%%%%%%%%%%%%%
%%%%% ** Reference ** %%%%%%%%%%%%%%%%%%%%%%%%%%%%%%%%%%%%%%%%%%%%%%%%
%%%%%%%%%%%%%%%%%%%%%%%%%%%%%%%%%%%%%%%%%%%%%%%%%%%%%%%%%%%%%%%%%%%%%%

%%%%%%%%%%%%%%%%%%%%%%%%%%%%%%%%%%%%%%%%%%%%%%%%%%%%%%%%%%%%%%%%%%%%%%%

%%%%%%%%%%%%%%%%%%%%%%%%%%%%%%%%%%%%%%%%%%%%%%%%%%%%%%%%%%%%%%%%%%%%%%%
%%%%%%%%%%%%%%%%%%%%%%%%%%%%%%%%%%%%%%%%%%%%%%%%%%%%%%%%%%%%%%%%%%%%%%%

\begin{thebibliography}{100}

\bibitem{Fukugita:1986hr}
  M.~Fukugita and T.~Yanagida,
  %``Baryogenesis Without Grand Unification,''
  Phys.\ Lett.\ B {\bf 174} (1986) 45.
  %doi:10.1016/0370-2693(86)91126-3

\bibitem{Buchmuller:2005eh}
  W.~Buchmuller, R.~D.~Peccei and T.~Yanagida,
  %``Leptogenesis as the origin of matter,''
  Ann.\ Rev.\ Nucl.\ Part.\ Sci.\  {\bf 55} (2005) 311
  %doi:10.1146/annurev.nucl.55.090704.151558
  [hep-ph/0502169].
  
\bibitem{Davidson:2008bu}
  S.~Davidson, E.~Nardi and Y.~Nir,
  %``Leptogenesis,''
  Phys.\ Rept.\  {\bf 466} (2008) 105
  %doi:10.1016/j.physrep.2008.06.002
  [arXiv:0802.2962 [hep-ph]].

%\cite{Kuzmin:1985mm}
\bibitem{Kuzmin:1985mm}
  V.~A.~Kuzmin, V.~A.~Rubakov and M.~E.~Shaposhnikov,
  %``On the Anomalous Electroweak Baryon Number Nonconservation in the Early Universe,''
  Phys.\ Lett.\  {\bf 155B} (1985) 36.
  %doi:10.1016/0370-2693(85)91028-7
  %%CITATION = doi:10.1016/0370-2693(85)91028-7;%%
  %2608 citations counted in INSPIRE as of 27 Dec 2018
  
%
\bibitem{Aghanim:2018eyx}
  N.~Aghanim {\it et al.} [Planck Collaboration],
  %``Planck 2018 results. VI. Cosmological parameters,''
  arXiv:1807.06209 [astro-ph.CO].


\bibitem{Giudice:2003jh}
	For example, see
  G.~F.~Giudice, A.~Notari, M.~Raidal, A.~Riotto and A.~Strumia,
  %``Towards a complete theory of thermal leptogenesis in the SM and MSSM,''
  Nucl.\ Phys.\ B {\bf 685} (2004) 89
  %doi:10.1016/j.nuclphysb.2004.02.019
  [hep-ph/0310123].

\bibitem{Leptogenesis_inflation_decay}
%\cite{Lazarides:1991wu}
%\bibitem{Lazarides:1991wu}
  G.~Lazarides and Q.~Shafi,
  %``Origin of matter in the inflationary cosmology,''
  Phys.\ Lett.\ B {\bf 258} (1991) 305;
  %%CITATION = PHLTA,B258,305;%%
  %216 citations counted in INSPIRE as of 23 May 2013


 
 %\cite{Asaka:1999yd}
%\bibitem{Asaka:1999yd}
  T.~Asaka, K.~Hamaguchi, M.~Kawasaki and T.~Yanagida,
  %``Leptogenesis in inflaton decay,''
  Phys.\ Lett.\ B {\bf 464} (1999) 12
  [hep-ph/9906366];
  %%CITATION = HEP-PH/9906366;%%
  %151 citations counted in INSPIRE as of 23 May 2013
%\cite{Asaka:1999jb}
%\bibitem{Asaka:1999jb}
%  T.~Asaka, K.~Hamaguchi, M.~Kawasaki and T.~Yanagida,
  %``Leptogenesis in inflationary universe,''
  Phys.\ Rev.\ D {\bf 61} (2000) 083512
  [hep-ph/9907559].
  %%CITATION = HEP-PH/9907559;%%
  %134 citations counted in INSPIRE as of 23 May 2013

\bibitem{Seesaw}
%\cite{Minkowski:1977sc}
%\bibitem{Minkowski:1977sc}
P.~Minkowski,
%``Mu $\to$ E Gamma At A Rate Of One Out Of 1-Billion Muon Decays?,''
Phys.\ Lett.\ B {\bf 67} (1977) 421;
%%CITATION = PHLTA,B67,421;%%
%%%%
T.~Yanagida,
in {\em Proc. of the Workshop on the Unified Theory
and the Baryon Number in the Universe}, 
Tsukuba, Japan, Feb.~13-14, 1979, p.~95, 
eds. O.~Sawada and S.~Sugamoto, 
(KEK Report KEK-79-18, 1979, Tsukuba); 
Progr.\ Theor.\ Phys.\ {\bf 64} (1980) 1103 ; 
%%%%
M.~Gell-Mann, P.~Ramond and R.~Slansky, 
in {\em Supergravity}, 
eds. P.~van~Niewenhuizen and D.~Z.~Freedman
(North Holland, Amsterdam 1980);
%%%%
P.~Ramond, 
in {\em Talk given at the Sanibel Symposium}, 
Palm Coast, Fla., Feb.~25-Mar.~2, 1979, preprint CALT-68-709
(retroprinted as hep-ph/9809459);
%%%%
S.~L.~Glashow,
in {\em Proc. of the Carg\'ese  Summer Institute on Quarks and Leptons},
Carg\'ese, July 9-29, 1979, 
eds. M.~L\'evy et. al, , (Plenum, 1980, New York), p707.

\bibitem{Pilaftsis:2003gt} 
  A.~Pilaftsis and T.~E.~J.~Underwood,
  %``Resonant leptogenesis,''
  Nucl.\ Phys.\ B {\bf 692}, 303 (2004)
  [hep-ph/0309342].
  %%CITATION = HEP-PH/0309342;%%
  %340 citations counted in INSPIRE as of 28 May 2013

\bibitem{Akhmedov:1998qx}
  E.~K.~Akhmedov, V.~A.~Rubakov and A.~Y.~Smirnov,
  %``Baryogenesis via neutrino oscillations,''
  Phys.\ Rev.\ Lett.\  {\bf 81} (1998) 1359.
%  [arXiv:hep-ph/9803255].
  

\bibitem{Asaka:2005pn} 
  T.~Asaka and M.~Shaposhnikov,
  %``The nuMSM, dark matter and baryon asymmetry of the universe,''
  Phys.\ Lett.\ B {\bf 620}, 17 (2005)
  [hep-ph/0505013].
  %%CITATION = HEP-PH/0505013;%%
  %286 citations counted in INSPIRE as of 16 Jul 2015


\bibitem{Canetti:2010aw}
  L.~Canetti and M.~Shaposhnikov,
  %``Baryon Asymmetry of the Universe in the NuMSM,''
  JCAP {\bf 1009} (2010) 001
  %doi:10.1088/1475-7516/2010/09/001
  [arXiv:1006.0133 [hep-ph]].

\bibitem{Asaka:2013jfa}
  T.~Asaka and S.~Eijima,
  %``Direct Search for Right-handed Neutrinos and Neutrinoless Double Beta Decay,''
  PTEP {\bf 2013} (2013) no.11,  113B02
  %doi:10.1093/ptep/ptt094
  [arXiv:1308.3550 [hep-ph]].

%\cite{Abada:2006fw}
\bibitem{Abada:2006fw}
  A.~Abada, S.~Davidson, F.~X.~Josse-Michaux, M.~Losada and A.~Riotto,
  JCAP {\bf 0604} (2006) 004
  %doi:10.1088/1475-7516/2006/04/004
  [hep-ph/0601083].

%\cite{Nardi:2006fx}
\bibitem{Nardi:2006fx}
  E.~Nardi, Y.~Nir, E.~Roulet and J.~Racker,
  JHEP {\bf 0601} (2006) 164
  %doi:10.1088/1126-6708/2006/01/164
  [hep-ph/0601084].
  
%\cite{Abada:2006ea}
\bibitem{Abada:2006ea}
  A.~Abada, S.~Davidson, A.~Ibarra, F.-X.~Josse-Michaux, M.~Losada and A.~Riotto,
  JHEP {\bf 0609} (2006) 010
  %doi:10.1088/1126-6708/2006/09/010
  [hep-ph/0605281].

%\cite{Blanchet:2006be}
\bibitem{Blanchet:2006be}
  S.~Blanchet and P.~Di Bari,
   JCAP {\bf 0703} (2007) 018
  %doi:10.1088/1475-7516/2007/03/018
  [hep-ph/0607330].

%\cite{Pascoli:2006ie}
\bibitem{Pascoli:2006ie}
  S.~Pascoli, S.~T.~Petcov and A.~Riotto,
  Phys.\ Rev.\ D {\bf 75} (2007) 083511
  %doi:10.1103/PhysRevD.75.083511
  [hep-ph/0609125].
 
 %\cite{Pascoli:2006ci}
\bibitem{Pascoli:2006ci}
  S.~Pascoli, S.~T.~Petcov and A.~Riotto,
  %``Leptogenesis and Low Energy CP Violation in Neutrino Physics,''
  Nucl.\ Phys.\ B {\bf 774} (2007) 1
 % doi:10.1016/j.nuclphysb.2007.02.019
  [hep-ph/0611338].
  %%CITATION = doi:10.1016/j.nuclphysb.2007.02.019;%%
  %169 citations counted in INSPIRE as of 20 Mar 2019
  
  %\cite{Moffat:2018smo}
\bibitem{Moffat:2018smo}
  K.~Moffat, S.~Pascoli, S.~T.~Petcov and J.~Turner,
  %``Leptogenesis from Low Energy $CP$ Violation,''
  JHEP {\bf 1903} (2019) 034
  doi:10.1007/JHEP03(2019)034
  [arXiv:1809.08251 [hep-ph]].
  %%CITATION = doi:10.1007/JHEP03(2019)034;%%
  %5 citations counted in INSPIRE as of 21 May 2019

 %\cite{DeSimone:2006nrs}
\bibitem{DeSimone:2006nrs}
  A.~De Simone and A.~Riotto,
  JCAP {\bf 0702} (2007) 005
  %doi:10.1088/1475-7516/2007/02/005
  [hep-ph/0611357].
     
\bibitem{PMNS}
%\bibitem{Pontecorvo:1958}
  B.~Pontecorvo, Sov.\ Phys.\ JETP\ {\bf 7} (1958) 172;\\
%\bibitem{Maki:1962mu}
  Z.~Maki, M.~Nakagawa and S.~Sakata,
  %``Remarks on the unified model of elementary particles,''
  Prog.\ Theor.\ Phys.\  {\bf 28} (1962) 870.
  %%CITATION = PTPKA,28,870;%%

\bibitem{Pas:2015eia}
	See, for example, a recent review, 
  H.~P\"as and W.~Rodejohann,
  %``Neutrinoless Double Beta Decay,''
  New J.\ Phys.\  {\bf 17} (2015) no.11,  115010
  %doi:10.1088/1367-2630/17/11/115010
  [arXiv:1507.00170 [hep-ph]].

\bibitem{Esteban:2016qun}
	NuFIT 3.2 (2018), www.nu-fit.org;
  I.~Esteban, M.~C.~Gonzalez-Garcia, M.~Maltoni, I.~Martinez-Soler and T.~Schwetz,
  %``Updated fit to three neutrino mixing: exploring the accelerator-reactor complementarity,''
  JHEP {\bf 1701} (2017) 087
  %doi:10.1007/JHEP01(2017)087
  [arXiv:1611.01514 [hep-ph]].

\bibitem{Casas:2001sr}
  J.~A.~Casas and A.~Ibarra,
  %``Oscillating neutrinos and mu --> e, gamma,''
  Nucl.\ Phys.\  B {\bf 618} (2001) 171
  [arXiv:hep-ph/0103065].
  %%CITATION = NUPHA,B618,171;%%

  %\cite{Garny:2011hg}
\bibitem{Garny:2011hg}
  M.~Garny, A.~Kartavtsev and A.~Hohenegger,
  %``Leptogenesis from first principles in the resonant regime,''
  Annals Phys.\  {\bf 328} (2013) 26
  %doi:10.1016/j.aop.2012.10.007
  [arXiv:1112.6428 [hep-ph]].
  %%CITATION = doi:10.1016/j.aop.2012.10.007;%%
  %81 citations counted in INSPIRE as of 30 Nov 2018
  
  %\cite{Iso:2013lba}
\bibitem{Iso:2013lba}
  S.~Iso, K.~Shimada and M.~Yamanaka,
  %``Kadanoff-Baym approach to the thermal resonant leptogenesis,''
  JHEP {\bf 1404} (2014) 062
  %doi:10.1007/JHEP04(2014)062
  [arXiv:1312.7680 [hep-ph]].
  %%CITATION = doi:10.1007/JHEP04(2014)062;%%
  %20 citations counted in INSPIRE as of 30 Nov 2018

%\cite{Dev:2014laa}
\bibitem{Dev:2014laa}
  P.~S.~Bhupal Dev, P.~Millington, A.~Pilaftsis and D.~Teresi,
  %``Flavour Covariant Transport Equations: an Application to Resonant Leptogenesis,''
  Nucl.\ Phys.\ B {\bf 886} (2014) 569
%  doi:10.1016/j.nuclphysb.2014.06.020
  [arXiv:1404.1003 [hep-ph]].
  %%CITATION = doi:10.1016/j.nuclphysb.2014.06.020;%%
  %111 citations counted in INSPIRE as of 25 Jan 2019
 
%\cite{Dev:2014wsa}
\bibitem{Dev:2014wsa}
  P.~S.~Bhupal Dev, P.~Millington, A.~Pilaftsis and D.~Teresi,
  %``Kadanoff–Baym approach to flavour mixing and oscillations in resonant leptogenesis,''
  Nucl.\ Phys.\ B {\bf 891} (2015) 128
%  doi:10.1016/j.nuclphysb.2014.12.003
  [arXiv:1410.6434 [hep-ph]].
  %%CITATION = doi:10.1016/j.nuclphysb.2014.12.003;%%
  %40 citations counted in INSPIRE as of 25 Jan 2019 

\bibitem{DOnofrio:2014rug}
  M.~D'Onofrio, K.~Rummukainen and A.~Tranberg,
  %``Sphaleron Rate in the Minimal Standard Model,''
  Phys.\ Rev.\ Lett.\  {\bf 113} (2014) no.14,  141602
  %doi:10.1103/PhysRevLett.113.141602
  [arXiv:1404.3565 [hep-ph]].
 
 %\cite{Bambhaniya:2016rbb}
\bibitem{Bambhaniya:2016rbb}
  G.~Bambhaniya, P.~S.~Bhupal Dev, S.~Goswami, S.~Khan and W.~Rodejohann,
  %``Naturalness, Vacuum Stability and Leptogenesis in the Minimal Seesaw Model,''
  Phys.\ Rev.\ D {\bf 95} (2017) no.9,  095016
%  doi:10.1103/PhysRevD.95.095016
  [arXiv:1611.03827 [hep-ph]].
  %%CITATION = doi:10.1103/PhysRevD.95.095016;%%
  %37 citations counted in INSPIRE as of 25 Jan 2019
  
\bibitem{NOvA:2018gge}
  M.~A.~Acero {\it et al.} [NOvA Collaboration],
  %``New constraints on oscillation parameters from $\nu_e$ appearance and $\nu_\mu$ disappearance in the NOvA experiment,''
  Phys.\ Rev.\ D {\bf 98} (2018) 032012
  %doi:10.1103/PhysRevD.98.032012
  [arXiv:1806.00096 [hep-ex]].

\bibitem{Abe:2018wpn}
  K.~Abe {\it et al.} [T2K Collaboration],
  %``Search for CP Violation in Neutrino and Antineutrino Oscillations by the T2K Experiment with $2.2\times10^{21}$ Protons on Target,''
  Phys.\ Rev.\ Lett.\  {\bf 121} (2018) no.17,  171802
  %doi:10.1103/PhysRevLett.121.171802
  [arXiv:1807.07891 [hep-ex]].

%\cite{Garbrecht:2014kda}
\bibitem{Garbrecht:2014kda}
  B.~Garbrecht and P.~Schwaller,
  %``Spectator Effects during Leptogenesis in the Strong Washout Regime,''
  JCAP {\bf 1410} (2014) no.10,  012
  %doi:10.1088/1475-7516/2014/10/012
  [arXiv:1404.2915 [hep-ph]].
  %%CITATION = doi:10.1088/1475-7516/2014/10/012;%%
  %13 citations counted in INSPIRE as of 29 Dec 2018  
  
\bibitem{Plumacher:1996kc}
  M.~Plumacher,
  %``Baryogenesis and lepton number violation,''
  Z.\ Phys.\ C {\bf 74} (1997) 549
  %doi:10.1007/s002880050418
  [hep-ph/9604229].

\bibitem{Plumacher:1998ex}
  %M.~Pl\"umacher,
  M.~Plumacher,
  %``Baryon asymmetry, neutrino mixing and supersymmetric SO(10) unification,''
  Ph.D. Thesis [hep-ph/9807557].
  %%CITATION = HEP-PH/9807557;%%
  %34 citations counted in INSPIRE as of 12 Jul 2018
 
     
 \end{thebibliography}
\end{document}